# Beyond Citations: Measuring Novel Scientific Ideas and their Impact in Publication Text

Sam Arts*
Department of management, strategy and innovation
Faculty of economics and business
KU Leuven
sam.arts@kuleuven.be

Nicola Melluso
Department of management, strategy and innovation
Faculty of economics and business
KU Leuven
nicola.melluso@kuleuven.be

Reinhilde Veugelers
Department of management, strategy and innovation
Faculty of economics and business
KU Leuven
reinhilde.veugelers@kuleuven.be

## ABSTRACT

New scientific ideas drive progress, yet measuring scientific novelty remains challenging. We use natural language processing to detect the origin and impact of new ideas in scientific publications. To validate our methods, we analyze Nobel Prize-winning papers, which likely pioneered impactful new ideas, and literature review papers, which typically consolidate existing knowledge. We also show that novel papers have more intellectual neighbors published after them, indicating they are ahead of their intellectual peers. Finally, papers introducing new ideas, particularly those with greater follow-on reuse, attract more citations.

* Corresponding author

Financial support from KU Leuven Grant 3H200208 is gratefully acknowledged, as are comments from participants at the 2023 NBER-SI-SSF workshop, the WOEPSR23 Conference, DRUID23, REGIS Summer School 2023 and 2024 OpenAlex Virtual User Conference, and particularly from Pierre Azoulay, Stefano Baruffaldi, Chiara Franzoni, Dietmar Harhoff, Vincent Thorge Holst, Jianan Hou, Erin Leahey, Rodrigo Ito and Fabio Montobbio.

# 1.      Introduction

New scientific ideas are fundamental to driving progress in science, technology, and economic prosperity (Bush, 1945; Mokyr, 2002). The significance of new scientific discoveries such as the transistor, magnetic resonance imaging, polymerase chain reaction, or carbon nanotubes cannot be overstated. However, despite their critical role, identifying and measuring novel scientific ideas—and tracing their diffusion and impact—remains challenging.

To measure new scientific ideas and the novelty of scientific papers, prior work has traditionally relied on their patterns of citations to prior work (Uzzi et al., 2013; Wang, Veugelers, & Stephan, 2017; Wu, Wang, & Evans, 2019). Yet, this approach has important limitations. Citations capture prior art but not the scientific content and contribution of the paper itself. Thus, citation-based methods may fail to accurately identify new scientific ideas and measure the novelty in the content of a paper (Fontana et al., 2020). Moreover, citations may not always reflect intellectual influence or impact of novel ideas. A survey of 9,000 academics across 15 fields found that more than half of cited papers had only minor or very minor intellectual influence (Teplitskiy et al., 2022).

In this paper, we move beyond citations and build on Kuhn's (1962) argument that scientific ideas are embedded in the text of scientific literature, making shifts in language critical for identifying new scientific ideas. We apply Natural Language Processing (NLP) techniques to harness the text of scientific publications, identifying both the origin and impact of new scientific ideas. By now, a substantial body of research, scattered across various disciplines from physics to sociology, has begun exploring NLP for detecting novel scientific ideas and their impact.[1] Yet, several shortcomings remain. First, existing work has yet to develop text metrics that can simultaneously identify new scientific ideas and measure a paper's novelty at the time of publication, as well as trace the diffusion and impact of these ideas over time.[2] Differentiating between novelty at publication and later adoption allows us

---

[1] For a comprehensive overview of papers utilizing text metrics, see Arts, Hou, and  Gomez (2021), Azoulay, Graff Zivin, and Manso (2011), Bhattacharya and Packalen (2020), Boudreau et al. (2016), Chai and Menon (2019), Cheng et al. (2023), Evans (2008, 2010), Fortunato et al. (2018), Foster, Rzhetsky, and Evans (2015), Gerow et al. (2018), Hofstra et al. (2020), Iaria, Schwarz, and Waldinger (2018), Kelly et al. (2021), Kuhn, Perc, and Helbing (2014), Milojević (2015), Packalen and Bhattacharya (2015, 2019, 2020), Park, Leahey, and Funk (2023), Shi and Evans (2023), Shibayama, Yin and Matsumoto (2021).

[2] The two papers most closely aligned with our approach are Cheng et al. (2023) and Hofstra et al. (2020). Cheng et al. (2023) track the diffusion of new scientific ideas using AutoPhrase, a machine learning tool trained on a broader corpus and sources such as Wikipedia, including post-publication data, to extract key phrases based on their prominence and repetition. This approach may inadvertently select scientific ideas that have already gained traction and proven successful. Furthermore, the authors only consider phrases that appear at least 120 times across scientific publications as novel ideas. As we illustrate

to separate the emergence of new ideas from their eventual success, facilitating an analysis of the factors influencing the success or failure of novel ideas. To accurately assess a paper's novelty, it is important to consider only prior work available at the time of publication, excluding any subsequent publications or data, to avoid success bias when identifying novel ideas. Second, existing studies lack comprehensive large-scale validation of text metrics. Without evidence that these metrics effectively capture new scientific ideas and their influence on future research, their utility remains questionable. Third, there is little consensus on which text metrics to adopt and whether they offer an improvement over traditional citation-based measures. The proliferation of proposed metrics complicates comparisons across studies, hindering scientific progress. Lastly, prior research has not provided open access to the underlying code, data, and metrics, which limits broader adoption. Given the scale of processing required for the entire corpus of scientific work, open access to these resources could facilitate the broader use of text metrics and support community efforts to further improve these measures. This paper aims to address these gaps in the literature.

We use the titles and abstracts of all scientific publications covered in the January 2024 snapshot of OpenAlex, which covers the largest corpus of scientific work published in the entire history from 1666 until 2023, and has the advantage of being open access (Priem, Piwowar, & Orr, 2022; Lin et al., 2023). To detect new scientific ideas and measure a paper's novelty at publication, we identify words, noun phrases, and novel combinations of words or noun phrases that appear for the first time. For example, we track the first papers introducing the word "*perceptron*," the noun phrase "*atomic force microscope*," or the word combination "*dna*" and "*microarray*." Alternatively, instead of distilling individual keywords, noun phrases, or combinations of these, we measure a paper's novelty based on the similarity of its entire text to all prior papers. To do this, we use SPECTER, a pre-trained document-level embedding of scientific papers, which has the advantage of accounting for synonyms, polysemy, and the context of words (Cohan et al., 2020).[3] To measure the impact or influence of new scientific ideas, we count the number of subsequent papers reusing new words, noun phrases, or combinations of either words or noun phrases. For instance, 19,495 papers reuse "*perceptron*," 17,322 reuse "*atomic force microscope*," and 25,916 reuse the combination of "*dna*" and "*microarray.*"

---

later, only the top 1.6% most impactful new noun phrases are reused at least 120 times. Hofstra et al. (2020) applies topic models to US doctoral dissertation titles and abstracts to assess novelty and how later dissertations incorporate these ideas. However, we believe that methods such as topic models and AutoPhrase may introduce bias by disproportionately identifying ideas that have already become significantly successful.

[3] Note that, like other machine learning tools (e.g., topic modeling and AutoPhrase), SPECTER is trained on the full corpus, including post-publication data, potentially biasing it toward identifying successful ideas.

To validate our text metrics and their improvement over traditional citation-based measures, we analyze Nobel Prize papers, which likely introduced impactful new ideas. The Nobel Prize also provides a means to validate whether the novel ideas identified by our text metrics across the entire corpus of scientific literature align with the contributions highlighted in the official Nobel Prize documentation. We also examine literature review papers, which typically consolidate existing knowledge rather than introduce novel ideas. Additionally, we show that more novel papers tend to have a higher proportion of their closest intellectual neighbors published after them, indicating they are ahead of their peers, and are more likely to use novelty-signaling language such as "discover" or "innovative." Finally, we demonstrate that papers introducing novel ideas—particularly those with substantial follow-on reuse— attract more citations and have a higher likelihood of becoming highly cited. These studies support the use of NLP in identifying new scientific ideas, measuring the novelty of papers at the time of publication, and assessing the impact of these ideas on subsequent scientific work. Furthermore, the results illustrate the improvement of text-based metrics over traditional citation-based measures. We provide open access to all data (https://zenodo.org/records/13869486) and code (https://github.com/nicolamelluso/science-novelty).

## 2. Identifying the Origin and Impact of New Scientific Ideas

### 2.1 Data collection

We collect all scientific publications from OpenAlex, concatenate the title and abstract of each paper, and process the text using standard NLP techniques: including tokenization, POS-tagging, dependency parsing, chunking, lemmatization, cleaning, baseline removal, and vectorization (see Appendix C for details).[4] We use title, abstract and full-text of publications published between 1666 and 1900 to construct a baseline dictionary and restrict the analysis to papers published between 1901 and 2023 (n=75,295,921).

### 2.2 Text metrics for novelty

First, we calculate *New Word* as the number of unique unigrams of a paper that appear for the first time in history. Hence, we identify the first paper introducing words such as "*apoptosis*" or "*photon*." Second, we calculate *New Phrase* as the number of unique noun phrases of a paper that appear for the first time in history.[5] Hence, we identify the first paper introducing noun phrases such as "*optical coherence tomography*" or "*vascular endothelial growth factor*." Third, we compute *New Word Combination* as the number of unique pairwise combinations of words that appear for the first time, regardless of the location or order of the words. For instance, we identify the first paper using the combination of words such as "*angiogenesis*" and "*therapeutic*" or "*carbon*" and *"nanotube.*" Fourth, we compute *New Phrase Combination* as the number of unique pairwise combinations of noun phrases that appear for the first time, regardless of the location or order of the phrases. For instance, we identify the first paper using the combination of phrases such as "*enzyme*" and "*superoxide dismutase*" or "*atom transfer*" and "*radical polymerization*." We exclude words, phrases, and combinations that appear only once in the entire corpus. Note that papers introducing new words or phrases introduce new combinations of words or phrases by construction. Finally, we generate a SPECTER document-embedding vector for each paper and calculate *Semantic Distance* as one minus the maximum cosine similarity between the focal paper and all prior papers from the past 5 years. This metric captures how distinct the focal paper is from the most similar prior work. For example, the seminal paper by Kirkpatrick, Gelatt, and Vecchi (1983) on simulated annealing scores in the top 1% for Semantic Distance.

---

[4] As 20% of the publications have only a title without an abstract, this could introduce potential bias. To address this, we include a binary indicator for abstract availability and control for text length in our regression analyses. Additionally, as a robustness check, we recalculated all text metrics using only titles across the full sample (see Appendix B). The main findings remain consistent, suggesting that titles alone can effectively identify new scientific ideas and their impact. However, the predictive power of title-based metrics is lower, highlighting the complementary value of abstracts.

### 2.3 Text metrics for impact of new scientific ideas

To assess the influence of new scientific ideas on subsequent research, we analyze how often these ideas appear in later publications. First, we calculate *New Word Reuse* as the number of new unigrams introduced by the focal paper, weighted by the number of subsequent papers that reuse these unigrams. For instance, 495,985 publications reuse "*apoptosis*" and 226,688 reuse "*photon*." For paper $p$, $New\ Word\ Reuse_p = \sum_{i=1}^{n}(1 + u_i)$ with $n$ representing the number of new unigrams introduced by paper $p$ and $u_i$ equal to the number of future papers which reuse the new unigram $i$. While *New Word* captures a paper's novelty at publication (ex ante), *New Word Reuse* measures the influence of the new scientific ideas on later research (ex post). Second, we compute *New Phrase Reuse* as the number of new noun phrases introduced by the focal paper, weighted by the number of later papers reusing those noun phrases. For example, "*optical coherence tomography*" is reused in 42,324 papers, and "*vascular endothelial growth factor*" in 55,897. Third, we compute *New Word Combination Reuse*, which measures the number of new word combinations weighted by the number of

---

later papers incorporating those combinations. For instance, 28,907 papers reuse the combination "*angiogenesis*" and "*therapeutic*" and 149,120 reuse the combination "*carbon*" and "*nanotube.*" Finally, we compute *New Phrase Combination Reuse* as the number of new noun phrase combinations weighted by the number of subsequent papers reusing those combinations. For instance, 6,518 papers reuse the combination of "*enzyme*" and "*superoxide dismutase*" and 6,144 papers reuse the combination of "*atom transfer*" and "*radical polymerization.*"

As shown in Appendix D, papers that reuse new words or noun phrases are approximately 53 and 45 times more likely, respectively, to cite the pioneering paper that introduced these words or phrases, compared to matched control papers from the same journal, year, and subfield that do not reuse these words or phrases. A similar pattern holds for the reuse of new word or phrase combinations. These findings suggest that the reuse of new ideas correlates with citations to the paper pioneering these ideas, confirming their intellectual influence. However, the data also reveal that only a small minority of reusing papers cite the pioneering paper, and as illustrated in Figure D.1, this likelihood declines significantly as the time between the pioneering and reusing paper increases. This suggests that researchers tend to cite more recent, closely related work rather than the foundational studies that originally introduced the insights upon which they build. These results highlight the complementary value of text-based metrics in capturing the diffusion and influence of new ideas, particularly in light of prior research indicating that citations alone provide a noisy and incomplete measure of intellectual influence (Cozzens, 1989; Teplitskiy et al., 2022).

### 2.4 Traditional metrics for novelty

Prior research traditionally relies on citations to measure the novelty of a paper. First, Uzzi et al. (2013) define novelty as an atypical combination of prior knowledge, where the observed frequency of any pair of cited journals is compared to the frequency of that pair occurring by chance. This comparison results in a normalized z-score that measures the (a)typicality of each pair of cited journals. The *Uzzi* score for a focal paper is determined as the $10^{th}$ percentile of the z-scores for all pairs of journals cited by the focal paper, i.e. focusing on the most atypical combinations introduced by the focal paper. Lower values of *Uzzi* indicate more atypical or novel papers.

Second, Wang, Veugelers and Stephan (2017) define a focal paper's novelty as the sum of the distances between first-time combinations of cited journals. Following their method, we first identify all pairs of journals cited together for the first time. We then calculate the distance for each new pair based on their co-citation frequencies and, finally, compute the *Wang* metric as the sum of these distances for all new combinations.

Finally, Funk and Owen-Smith (2017) introduce a metric (*CD*) that characterizes whether a paper is disruptive or consolidating. A paper is considered disruptive if it is cited without its predecessors being cited—those papers that it references. Conversely, a paper is seen as consolidating if it is frequently cited alongside its predecessors. *CD* cannot measure a paper's novelty at the time of publication (ex-ante) since it relies on subsequent citations (ex-post). Unlike *Uzzi* and *Wang*, *CD* is not a direct measure of a paper's novelty but rather a measure of the nature of its impact. Nevertheless, we include this measure in our analysis due to its growing use in the science of science community (e.g., Wu, Wang, & Evans, 2019; Park, Leahey, & Funk, 2023). In our analysis, we use the *Uzzi* atypicality and *CD* scores provided by SciSciNet (Lin et al., 2023).[6]

In this paper, we primarily use the novelty metrics (both citation-based and text-based) as continuous or count variables. Given the skewness of these measures, we also performed robustness checks by transforming them into binary indicators, classifying papers as novel if they rank in the top 5% (or top 1%) of a given metric within the same scientific subfield and year of publication. Although not reported, these checks confirmed that our results are robust across all validation tests, regardless of whether we use the raw measures or their binary transformations.

## 3. Descriptive Statistics

Table A.1 presents summary statistics at the level of new words, noun phrases, and their pairwise combinations. Between 1901 and 2023, approximately 6 million new words, 27 million new noun phrases, 1 billion new word combinations, and about 0.8 billion new noun phrase combinations were introduced. As expected, most new words, phrases, and combinations are reused by a small number of papers, while a select few are widely adopted. Among the most influential are the word "*positron*," first introduced in 1933 and reused by 96,988 publications; the noun phrase "*x-ray diffraction*," introduced in 1914 and reused by 344,357 papers; the combination of words "*electron*" and "*microscopy*," appearing in 1934 and reused by 674,072 papers; and the combination of noun phrases "*catalase*" and

---

"*superoxide dismutase*," originating in 1970 and reused in 35,423 papers. Table A.2 presents the top 10 most frequently reused new noun phrases introduced in each decade.

Table 1 presents summary statistics at the paper level, with Panel A showing ex ante metrics, which measure novelty at the time of publication, and Panel B showing ex post metrics, which assess impact after publication. Only a subset of papers introduce new scientific ideas. Approximately 7% of papers introduce a new word, 25% introduce a new noun phrase, 61% introduce a new word combination, and 65% introduce a new noun phrase combination. The average (median) paper introduces 0.08 (0) new words, 0.36 (0) new noun phrases, 13.22 (2) new word combinations, and 10.98 (2) new noun phrase combinations. Although the number of new word and noun phrase combinations may appear substantial, it is important to note that the average paper contains 816 unique word pairs and 157 unique noun phrase pairs. Consequently, the average (median) share of new word combinations per paper is only 1.9% (0.4%), and the average share of new noun phrase combinations is 9.0% (4.5%). As expected, the summary statistics reveal a significant skew across all text-based novelty metrics, and underscore the need to control for text length when assessing a paper's novelty.

'Table 1'

Figure 1 illustrates the average number of new phrases introduced by papers across all fields of study from 1901 to 2023, revealing significant variation in scientific novelty both across fields and over time.[7] For example, fields like Physics and Astronomy and Material Science exhibit higher rates of scientific novelty compared to others. The time trends also reflect the historical evolution of scientific fields, such as the surge in groundbreaking discoveries in Biochemistry, Genetics, and Molecular Biology after 1950. As shown in Figure A.1 , this substantial heterogeneity in scientific novelty persists even at the more granular subfield level.

'Figure 1'

The novelty of scientific papers varies both across subfields and within the same subfield over time and among papers published in the same subfield and year. A variance decomposition model using *New Phrase* as a novelty measure reveals that differences between subfields explain approximately 2% of the total variance, while variation within subfields over time accounts for 5%. The remaining 93% is attributed to differences among papers within the same subfield and year. Certain subfields and time periods, such as Molecular Biology from 1974 to 1976, stand out for particularly high rates of novel scientific ideas.

---

[7] OpenAlex assigns each paper to one of 252 subfields within a hierarchy of 26 fields, using subfield labels derived from Scopus. A machine learning model determines the appropriate subfield for each paper based on the text of its title and abstract, as well as co-citation patterns. More information can be found here: https://docs.openalex.org/api-entities/topics and a list of subfields can be found here https://api.openalex.org/topics.

The novelty of papers also varies significantly across journals. Differences between journals account for about 10% of the variance, variation within journals over time contributes 7%, and differences among papers within the same journal and year explain the remaining 83%. Journals like Cell and the Journal of Experimental Medicine consistently publish a notable share of papers introducing new scientific ideas. Interestingly, the relatively modest contribution of temporal variation is somewhat surprising given previous findings on the declining novelty (or disruptiveness) of papers over time (Bloom et al., 2020; Park, Leahey, & Funk, 2023).

As shown in Table A.3 of the Appendix, there is a strong positive correlation among text-based novelty metrics, indicating that novel ideas in a paper are often captured by multiple metrics simultaneously. However, average correlations between text-based and traditional citation-based novelty metrics are low, suggesting that text metrics assess novelty in a fundamentally different way than citation-based measures.

Finally, we examine the paper-level summary statistics for metrics that capture the impact of new scientific ideas. As shown in Panel B of Table 1, the average (median) value for *New Word Reuse* is 3.4 (0), for *New Phrase Reuse* 6.6 (0), for *New Word Combination Reuse* 197.8 (6), and for *New Phrase Combination Reuse* 67.8 (6). These descriptive statistics again highlight the significant skew in the reuse of new scientific ideas, with only a small minority of novel papers having a substantial impact.

## 4.     Nobel Prize Papers

Nobel Prize papers are recognized for pioneering new scientific ideas, particularly those with a profound impact on scientific progress. As such, they provide a benchmark for validating the effectiveness of our text metrics in identifying novel scientific ideas and their influence on subsequent scientific work. Using official documentation of laureates' contributions, we evaluate whether the novel scientific ideas identified by our metrics in Nobel Prize papers align with those described in the official documentation. Additionally, we perform a case-control study to evaluate the effectiveness of text-based and citation-based metrics in distinguishing Nobel Prize papers from matched control papers, based on the assumption that Nobel Prize papers are more likely to introduce new scientific ideas, particularly those with a significant impact on subsequent research.

*4.1 Data*

We collect papers linked to Nobel prizes in Chemistry, Physics and Physiology or Medicine from Li et al. (2019). Each Nobel prize paper is randomly matched to one control paper not linked to any Nobel prize but published in the same journal, year and subfield.[8] Our sample includes 584 Nobel Prize papers published between 1902 and 2007, linked to 234 Nobel Prizes awarded between 1908 and 2016. Among the 234 Nobel Prizes, at least one corresponding paper introduces a new word for 43%, a new phrase for 75%, a new word combination for 79%, and a new phrase combination for 85%.

For each Nobel Prize, we collected official online documentation detailing the recognized contributions, including the summary page, press release, and Nobel Lecture(s). Nobel Lectures are authored by the laureates on topics relevant to the work for which the prize was awarded (Nobel Foundation, 2024).[9] The texts were combined, preprocessed similarly to the papers, and references to the titles of Nobel Prize papers were removed (see Appendix C). Reassuringly, 88% of new words, 92% of new noun phrases, 90% of new word combinations, and 91% of new noun phrase combinations introduced by Nobel Prize papers are also mentioned in the corresponding official documentation. This demonstrates that our text metrics are effective, though not perfect, in identifying new scientific ideas across the entire body of scientific literature.

'Table 2'

Table 2 presents examples of Nobel Prizes, highlighting the prize motivation, a corresponding paper, and a new scientific idea identified by our text metrics that aligns with the Nobel Prize motivation. For instance, William Shockley, John Bardeen, and Walter Brattain shared the 1956 Nobel Prize in Physics "for their research on semiconductors and the discovery of the transistor effect." One of their corresponding papers, published in Physical Review in 1948, was the first to introduce the term "*transistor*," which has since been reused by 128,821 subsequent papers. In 1997, Stanley Prusiner was awarded the Nobel Prize in Medicine "for his discovery of prions–a new biological principle of infection." His corresponding paper, published in Science in 1982, introduced the combination of the words "*prion*" and "*protein*," which has been reused by 13,779 papers. The 1993 Nobel Prize in Chemistry was awarded to Kary Mullis "for his invention of the polymerase chain reaction method." His corresponding paper, published in 1986 in Cold Spring Harbor Symposia on Quantitative Biology, pioneered the noun phrase "*polymerase chain reaction*," which has been reused by 116,146 papers. Lastly, in 2011, Ralph M. Steinman received the Nobel Prize in Medicine "for his discovery of the dendritic cell and its role in adaptive immunity." His corresponding paper, published in 1973 in the Journal of Experimental Medicine, was the first to introduce the combination of the phrases "*dendritic cell*" and "*macrophage*," which has since been reused by 8,299 papers.

---

[8] We use 252 subfield classifications from OpenAlex. If no subfield match is available, we use the coarser 26 field classifications.
[9] Papers in our sample are explicitly referenced in these lectures (Lin et al., 2019). We collected a total of 481 lectures, with each laureate delivering one and each prize averaging two lectures. Of these, 91% were available as PDFs, while the remaining were plain text sourced directly from the laureates' Nobel Prize pages.

### 4.2 Descriptive results

Table A.4 presents descriptive statistics for Nobel Prize papers and matched control papers. Both text-based and traditional citation-based metrics effectively distinguish Nobel Prize papers from control papers, as indicated by the significant results of the Mann-Whitney test (p=0.000). The only exceptions are *Semantic Distance* and *Uzzi*. All text metrics that measure novelty at the time of publication (*New Word*, *New Phrase*, *New Word Combination*, *New Phrase Combination*) outperform traditional citation-based novelty measures (*Wang* and *Uzzi*). These findings provide evidence that text metrics are more effective at identifying new scientific ideas and measuring the novelty of papers at the time of publication. Among the text metrics, *New Phrase* performs best in identifying Nobel Prize papers.

As expected, and illustrated in Panel B, text metrics that measure the reuse of new scientific ideas after publication (*New Word Reuse*, *New Phrase Reuse*, *New Word Combination Reuse*, *New Phrase Combination Reuse*) also distinguish Nobel Prize papers from control papers, outperforming all metrics that capture novelty at the time of publication. Each of these impact-based text metrics also outperforms *CD*, the disruptiveness index. Of all metrics, *New Phrase Reuse* performs the best in identifying Nobel Prize papers. As Nobel Prize papers are renowned for pioneering impactful ideas, these findings demonstrate that text metrics tracking idea reuse effectively capture the influence and lasting impact of their novel contributions.

### 4.3 Regressions

Table 3 presents results from logit regressions with a binary indicator for Nobel Prize paper as the outcome variable. The models control for the number of unique words and phrases in the title and abstract of the paper, as well as whether the paper has an abstract available (Letchford, Moat & Preis, 2015; Milojević, 2017). Additionally, we account for the number of cited papers and journals and include fixed effects for publication year and subfield. To evaluate the performance of different metrics in correctly classifying Nobel Prize and control papers, we calculate precision (proportion of correctly classified Nobel Prize papers), recall (proportion of actual Nobel Prize papers correctly identified), and the area under the curve (AUC), which ranges from 0.5 (no predictive power) to 1 (perfect classification). Average marginal effects quantify the increase in the likelihood of a paper being a Nobel Prize paper per one standard deviation increase in each metric.

'Table 3'

The regression results in Table 3 align closely with the descriptive statistics in Table A.4. Panel A shows that all text metrics measuring novelty at the time of publication are significant at the 1% level and outperform traditional citation-based metrics, except *for Semantic Distance*, which is not significant at conventional levels. Binary versions of text metrics, such as *New Word (Binary)*, which identifies whether a paper introduced at least one new word, are statistically significant but generally perform worse and are omitted from the table for brevity. This indicates that Nobel Prize papers typically introduce multiple new words, noun phrases, or combinations, making count-based measures more effective at capturing novelty than binary indicators. Consistent with prior literature, traditional citation-based metrics (*Wang* and *Uzzi*) fail to distinguish Nobel Prize papers from controls (Fontana et al., 2020). Of all novelty metrics, *New Phrase* has the strongest discriminatory power.

In Model 8, which includes all text metrics, *New Word* becomes statistically insignificant, but this model achieves higher precision, recall, and predictive power compared to using *New Phrase* alone, the best-performing single metric. In Model 9, which incorporates both text and citation-based metrics, predictive power remains unchanged. These results highlight the effectiveness of text metrics, particularly *New Phrase*, as indicators of a paper's novelty at the time of publication.

Panel B shows that text metrics capturing the reuse of new scientific ideas generally outperform those measuring novelty at the time of publication in predicting Nobel Prizes. This finding is reassuring, as it suggests that text metrics effectively capture the broader influence and enduring impact of novel contributions introduced by Nobel Prize papers. Among all metrics, *New Phrase Reuse* emerges as the strongest predictor.

While the text metrics demonstrate reasonable accuracy in predicting Nobel Prize papers, they are obviously far from perfect. On one hand, a notable portion of predicted Nobel Prize papers are actually control papers, i.e., false positives. However, these control papers may also introduce new scientific ideas but simply did not receive a Nobel Prize, making them not necessarily false positives for novelty. For instance, a paper by Federico Capasso and colleagues published in Science in 2000 (in the same year, journal, and subfield as John Hall's Nobel-winning work) introduced the phrase "*midinfrared quantum cascade laser.*" On the other hand, some Nobel Prize papers are missed by the text metrics, i.e., false negatives. An example is the set of papers associated with the 2004 Nobel Prize in Chemistry awarded to Aaron Ciechanover, Avram Hershko, and Irwin Rose for their discovery of ubiquitin-mediated protein degradation. Additionally, while Nobel Prize-winning papers provide a useful validation benchmark, they have inherent limitations. The sample size is small, the focus is limited to certain fields, and novelty alone is not the sole criterion for awarding a Nobel Prize. These papers primarily reflect highly impactful new ideas rather than a complete representation of novel contributions in science.

## 5.    Literature Review Papers

As a second validation, we collect a large sample of literature review papers (n=34,428) along with a matched control sample of original, non-review papers. Our assumption is that review papers primarily summarize existing scientific knowledge rather than introducing new scientific ideas (Wu, Wang, & Evans, 2019). In other words, we expect review papers to be less likely to pioneer new insights compared to control papers. As detailed in Appendix E, the data and results support this assumption and align closely with the Nobel Prize validation findings. Given that the vast majority of scientific papers are perhaps not very novel, it is unsurprising that the text metrics are less effective at distinguishing review papers from control papers than at differentiating Nobel Prize papers from controls.

## 6.    Publication Timing of Intellectual Neighbors

As an additional validation, we check whether more novel papers have a greater proportion of their closest intellectual neighbors—defined as the most similar papers—published after them, compared to less novel papers. In other words, when a paper introduces a new scientific idea, related research is expected to be published after it, indicating that the paper is ahead of its intellectual peers. To test this, we identify all OpenAlex papers indexed in PubMed and published between 1901 and 2010 (n=11,542,812) and use the PubMed Related Citations Algorithm to find the five most similar articles for each paper based on title, abstract, and MeSH terms (Lin & Wilbur, 2007). As outlined in Appendix F, we find support for this using both text-based and citation-based novelty metrics, though the text-based metrics generally show a stronger effect on the proportion of intellectual neighbors published after the focal paper.

## 7.  Language Denoting Novelty

Papers introducing new scientific ideas are more likely to use language explicitly signaling novelty, such as terms like "discover," "introduce," "novel," or "innovative." Following the approach of Leahey et al. (2023), we construct a binary indicator for papers that include any of these novelty-signaling terms in their title or abstract and test the ability of all metrics to correctly classify such papers. This analysis is conducted on the full sample of papers published between 1901 and 2023 (n=75,295,921). As illustrated in Appendix G, while all metrics demonstrate some predictive power, text-based metrics generally outperform others in identifying papers that explicitly highlight their novelty through this type of language.

## 8.  New Scientific Ideas Fuel Scientific Progress

The motivation to identify papers that pioneer novel scientific ideas stems from their potential to significantly advance scientific progress, traditionally measured by the citations these papers receive (e.g., Uzzi et al., 2013; Wang, Veugelers, & Stephan, 2017). There are persistent concerns about the science funding system's ability to adequately support novel research, potentially resulting in missed opportunities for scientific progress and breakthroughs (Azoulay, Graff Zivin, & Manso, 2011; Alberts et al., 2014; Franzoni, Stephan, & Veugelers, 2022). In Appendix H, we analyze the full sample of papers published between 1901 and 2010 (n=37,154,406) and show that papers introducing new scientific ideas at the time of publication tend to attract more citations over time and are more likely to become highly cited. Figure 2 illustrates how the likelihood of a paper being among the top 1% most-cited (above the 99th percentile in citations within the same subfield and year) varies across percentile ranges of each novelty metric. Notably, text-based metrics capturing a paper's novelty at publication generally predict top-cited papers more effectively than traditional novelty metrics (*Uzzi* and *Wang*). Reassuringly, text-based metrics capturing the reuse of new scientific ideas generally outperform those measuring novelty at the time of publication in predicting citation outcomes. This indicates that, despite our earlier finding that papers reusing new scientific ideas do not consistently cite the pioneering papers introducing these ideas, these metrics effectively capture the broader influence and impact of new scientific ideas on subsequent literature, as reflected in citation counts.

'Figure 2'

## 9.    Discussion and Conclusion

New scientific insights drive progress in science, technology, and the economy. However, identifying and measuring novel scientific ideas—and tracing their diffusion and impact—remains a persistent challenge. Citation-based metrics have traditionally served as proxies for novelty—or, relatedly, atypicality or disruptiveness—in scientific papers. Yet, citations primarily reflect connections to prior work rather than the intrinsic scientific content or contributions of the papers themselves. Moreover, while citations are widely accepted as a measure of impact, they often serve rhetorical or strategic purposes, such as lending authority or adhering to conventions, rather than signaling substantive intellectual influence (Cozzens, 1989; Teplitskiy et al., 2022). As a result, citation-based metrics often fall short in accurately identifying novel scientific ideas at the time of publication and in capturing their true intellectual impact on scientific progress (Fontana et al., 2020).

In this paper, we employ natural language processing techniques to analyze and vectorize scientific content, enabling the identification of novel ideas and the measurement of scientific novelty at the time of publication, as well as the assessment of their subsequent reuse in the literature. Our findings confirm the effectiveness of text-based metrics in identifying novelty and measuring its impact, demonstrating that these metrics outperform traditional citation-based approaches.

Interestingly, we find that papers reusing new scientific ideas do not consistently cite the original work that introduced the idea, particularly when the idea was established long ago. This highlights how text-based metrics offer a novel perspective on the diffusion and influence of new ideas, extending beyond citation-based measures.

However, text-based approaches are not without limitations (particularly those stemming from the quality of underlying publication data) which future researchers must carefully consider. Appendix I provides a detailed overview of these limitations and offers suggestions for advancing this line of research. Despite these challenges, a key contribution of text-based metrics is their ability to capture the two main stages of scientific progress: the discovery of new ideas and their subsequent diffusion and use. This capability opens new avenues for studying not only the emergence of scientific ideas but also their influence and spread across the scientific community.

## Figure 1: Scientific Novelty by Field of Study and Year

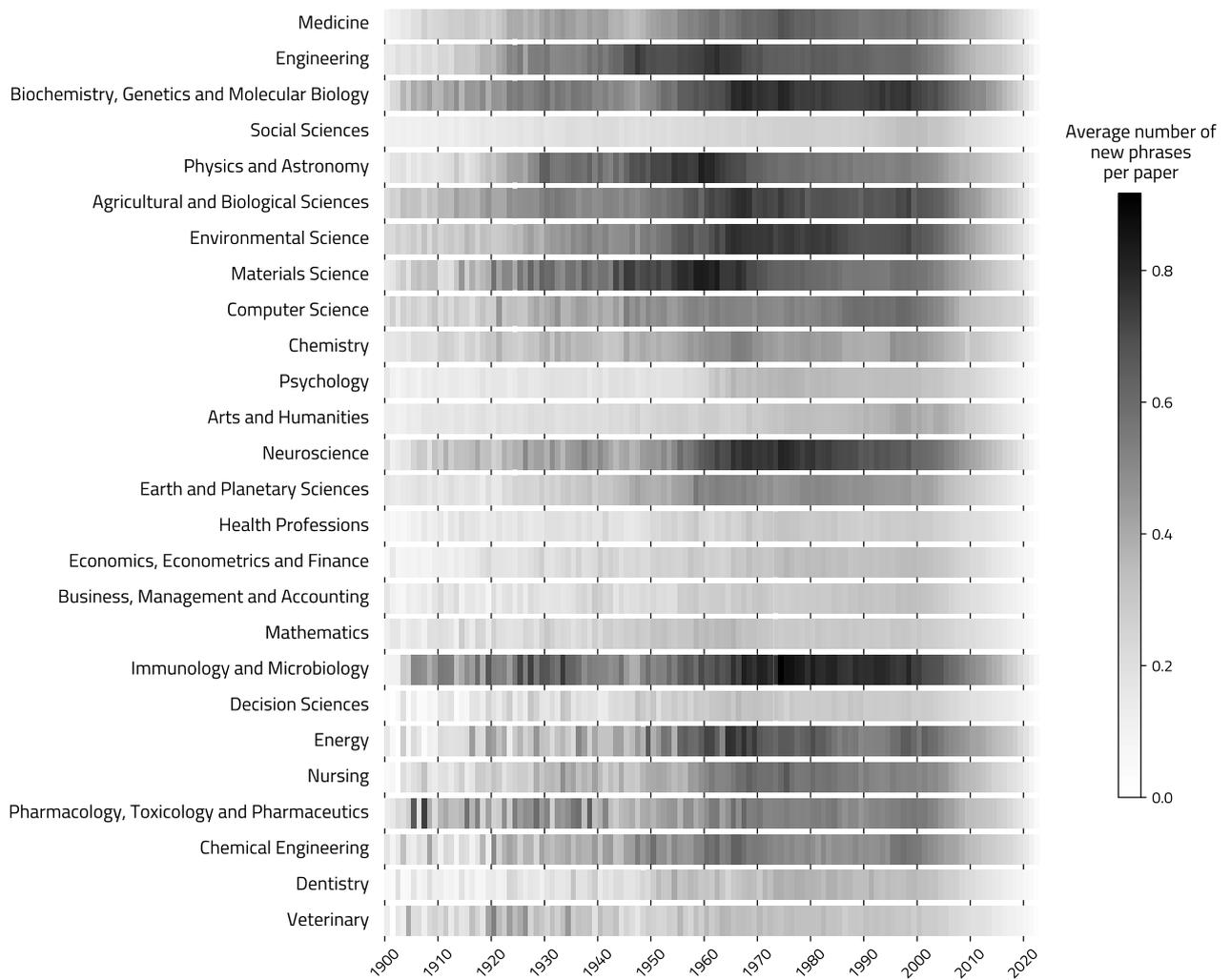

*Notes*: Paper-level average number of new noun phrases by fields of study (OpenAlex classification) and year (n= 75,295,921 papers published between 1901 and 2023). Fields are top down ordered by the overall number of publications.

## Figure 2: Predicted Probability of Paper Being Top Cited

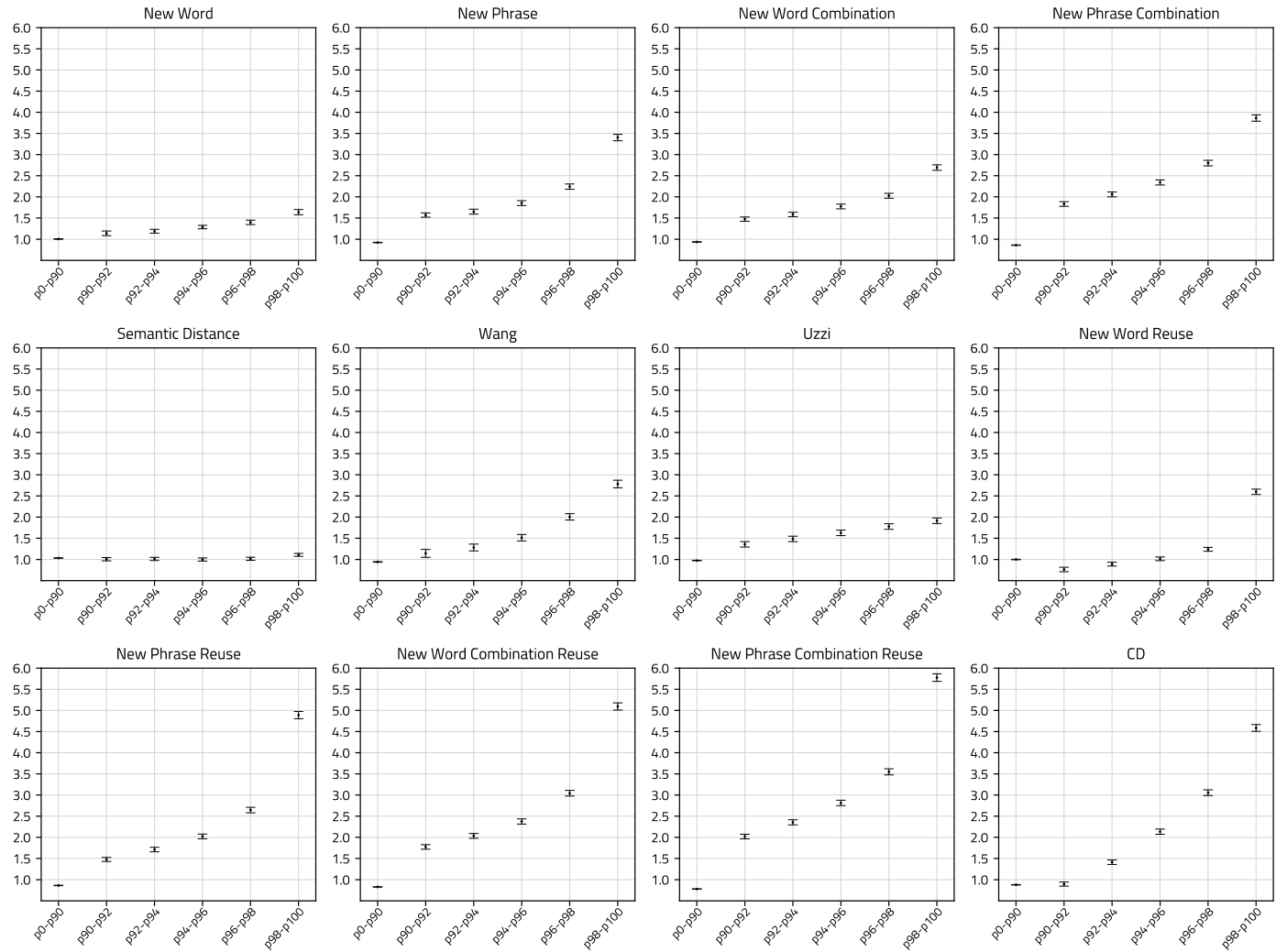

*Notes:* Figure 2 plots the predicted probability (in %) of a paper being among the top 1% most-cited (above the 99[th] percentile of received citations within its subfield and year), estimated using 12 separate linear probability models. Each model includes indicator variables for whether a paper's metric value, within its subfield and year, falls into the 0[th]–90[th] percentile (reference category) or one of five 2-percentile ranges from the 90[th] to 100[th] percentile. For consistency in comparisons, *Uzzi* is inverted (e.g., p98–p100 corresponds to p0–p2), ensuring higher novelty aligns with higher percentile values. The predicted probabilities are shown with 99.999% confidence intervals. Control variables in the models include whether the paper has an abstract, the number of unique words and phrases in title and abstract, the number of unique cited papers and journals, and fixed effects for subfield and year of publication. The figure is organized into 12 panels, each representing a different metric. The analysis includes all papers from the OpenAlex database published between 1901 and 2010 (n=37,154,406).

# Table 1: Summary Statistics Paper Level

| | Mean | St.Dev. | Min | p25 | p50 | p75 | p95 | p99 | Max | Skew |
|---|---|---|---|---|---|---|---|---|---|---|
| **Panel A: Ex ante** | | | | | | | | | | |
| New Word (Binary) | 0.068 | 0.252 | 0 | 0 | 0 | 0 | 1 | 1 | 1 | 3.432 |
| New Word | 0.083 | 0.361 | 0 | 0 | 0 | 0 | 1 | 2 | 281 | 17.244 |
| New Phrase (Binary) | 0.245 | 0.430 | 0 | 0 | 0 | 0 | 1 | 1 | 1 | 1.184 |
| New Phrase | 0.360 | 0.779 | 0 | 0 | 0 | 0 | 2 | 3 | 223 | 4.128 |
| New Word Combination (Binary) | 0.614 | 0.487 | 0 | 0 | 1 | 1 | 1 | 1 | 1 | -0.469 |
| New Word Combination | 13.222 | 97.659 | 0 | 0 | 2 | 11 | 55 | 147 | 256,995 | 980.874 |
| New Phrase Combination (Binary) | 0.649 | 0.477 | 0 | 0 | 1 | 1 | 1 | 1 | 1 | -0.624 |
| New Phrase Combination | 10.948 | 45.742 | 0 | 0 | 2 | 11 | 47 | 104 | 44,871 | 139.806 |
| Semantic Distance | 0.136 | 0.046 | 0.010 | 0.105 | 0.131 | 0.162 | 0.220 | 0.271 | 0.740 | 0.700 |
| Wang | 0.192 | 2.088 | 0 | 0 | 0 | 0 | 0.844 | 4.208 | 6,052.686 | 503.870 |
| Uzzi | 14.911 | 147.781 | -445.270 | 0 | 0 | 0.034 | 51.951 | 363.955 | 67,782 | 43.508 |
| **Panel B: Ex post** | | | | | | | | | | |
| New Word Reuse | 3.368 | 367.690 | 0 | 0 | 0 | 0 | 2 | 20 | 769,278 | 812.412 |
| New Phrase Reuse | 6.555 | 268.222 | 0 | 0 | 0 | 0 | 12 | 71 | 416,735 | 522.944 |
| New Word Combination Reuse | 197.827 | 2,322.053 | 0 | 0 | 6 | 52 | 584 | 3,061 | 1,359,901 | 119.531 |
| New Phrase Combination Reuse | 67.818 | 402.917 | 0 | 0 | 6 | 40 | 271 | 932 | 275,329 | 79.233 |
| CD | 0.002 | 0.039 | -1 | 0 | 0 | 0 | 0.005 | 0.094 | 1 | 9.578 |
| **Panel C: Controls** | | | | | | | | | | |
| Abstract (Binary) | 0.801 | 0.400 | 0 | 1 | 1 | 1 | 1 | 1 | 1 | -1.504 |
| N. of Words | 31.767 | 25.579 | 1 | 9 | 30 | 47 | 71 | 103 | 2,489 | 3.089 |
| N. of Phrases | 13.373 | 12.215 | 0 | 3 | 11 | 20 | 34 | 48 | 1,728 | 3.249 |
| N. of cited Papers | 18.203 | 26.856 | 0 | 0 | 9 | 28 | 62 | 114 | 6,315 | 6.304 |
| N. of cited Journals | 9.611 | 12.774 | 0 | 0 | 5 | 15 | 33 | 54 | 1,132 | 2.735 |

*Notes*: n= 75,295,921 papers published between 1901 and 2023. p25, p50, p75, p95 and p99 are respectively the 25[th], 50[th], 75[th], 95[th] and the 99[th] percentile. The skewness (skew) of the distributions is the measure of the asymmetry of the probability distribution of a real-valued random variable about its mean. A positive skewness indicates that the distribution has a long right tail, while a negative skewness indicates a long left tail.

## Table 2: Examples of Nobel Prizes

| Prize | Short Prize Motivation | Paper | New word (phrase) or new combination of words (phrases) |
|---|---|---|---|
| Chemistry 1934 | Discovery of heavy hydrogen | Urey, Harold C., Ferdinand G. Brickwedde, and George M. Murphy, "A Hydrogen Isotope of Mass 2," *Physical Review* 39:1 (1932), 164. | hydrogen_isotope (4,074) |
| Physics 1936 | Discovery of the positron | Anderson, Carl D., "The Positive Electron," *Physical Review* 43:6 (1933), 491. | positron (94,146) |
| Medicine 1952 | Discovery of streptomycin, the first antibiotic effective against tuberculosis | Schatz, Albert, Elizabeth Bugle, and Selman A. Waksman, "Streptomycin, a Substance Exhibiting Antibiotic Activity Against Gram-Positive and Gram-Negative Bacteria," *Proceedings of the Society for Experimental Biology and Medicine* 55:1 (1944), 66–69. | streptomycin (23,334) |
| Physics 1956 | Researches on semiconductors and discovery of the transistor effect | Bardeen, John, and Walter H. Brattain, "The Transistor, a Semi-Conductor Triode," *Physical Review* 74:2 (1948), 230. | transistor (152,047) |
| Physics 1959 | Discovery of the antiproton | Chamberlain, Owen, Emilio Segrè, Clyde Wiegand, and Thomas Ypsilantis, "Observation of Antiprotons," *Physical Review* 100:3 (1955), 947. | antiproton (5,023) |
| Medicine 1962 | Discoveries concerning the molecular structure of nucleic acids and its significance for information transfer in living material | Watson, James D., and Francis H. C. Crick, "The Structure of DNA," *Cold Spring Harbor Symposia on Quantitative Biology* 18:1 (1953), 123–131. | dna,genetic_material (1,985) |
| Medicine 1963 | Discoveries concerning the ionic mechanisms involved in excitation and inhibition in the peripheral and central portions of the nerve cell membrane | Eccles, John C., Peter Fatt, and Koichi Koketsu, "Cholinergic and Inhibitory Synapses in a Pathway from Motor-Axon Collaterals to Motoneurones," *The Journal of Physiology* 126:3 (1954), 524. | inhibitory_synapsis (2,275) |
| Medicine 1969 | Discoveries concerning the replication mechanism and the genetic structure of viruses | Hershey, Alfred D., and Martha Chase, "Independent Functions of Viral Protein and Nucleic Acid in Growth of Bacteriophage," *The Journal of General Physiology* 36:1 (1952), 39–56. | dna,viral (58,844) |
| Physics 1971 | Discovery for the invention and development of the holographic method | Gabor, Dennis, "A New Microscopic Principle," *Nature* 161:4098 (1948), 777–778. | holography (6,829) |
| Medicine 1979 | Development of computer assisted tomography | Hounsfield, Godfrey N., "Computerized Transverse Axial Scanning (Tomography): Part 1. Description of System," *The British Journal of Radiology* 46:552 (1973), 1016–1022. | computerize,tomography (21,835) |
| Physics 1986 | Fundamental work in electron optics, and for the design of the first electron microscope | Binnig, Gerd, Heinrich Rohrer, Christoph Gerber, and Ed Weibel, "Surface Studies by Scanning Tunneling Microscopy," *Physical Review Letters* 49:1 (1982), 57. | scanning_tunnel_microscope (17,914) |
| Chemistry 1993 | Invention of the polymerase chain reaction (PCR) method | Mullis, Kary, Fred Faloona, Sherry Scharf, Randall Saiki, Glenn Horn, and Henry Erlich, "Specific Enzymatic Amplification of DNA In Vitro: The Polymerase Chain Reaction," *Cold Spring Harbor Symposia on Quantitative Biology* 51:1 (1986), 263–273. | polymerase_chain_reaction (209,968) |
| Medicine 1997 | Discovery of Prions - a new biological principle of infection | Prusiner, Stanley B., "Novel Proteinaceous Infectious Particles Cause Scrapie," *Science* 216:4542 (1982), 136–144. | prion,protein (13,779) |
| Medicine 1998 | Discoveries concerning nitric oxide as a signalling molecule in the cardiovascular system | Katsuki, Shiro, Walter Arnold, Charanjit Mittal, and Ferid Murad, "Stimulation of Guanylate Cyclase by Sodium Nitroprusside, Nitroglycerin, and Nitric Oxide in Various Tissue Preparations and Comparison to the Effects of Sodium Azide and Hydroxylamine," *Journal of Cyclic Nucleotide Research* 3:1 (1977), 23–35. | nitric_oxide,stimulation (4,641) |
| Chemistry 2001 | Work on chirally catalysed hydrogenation reactions | Nozaki, Hiroshi, Hiroshi Takaya, Seiichi Moriuti, and Ryoji Noyori, "Homogeneous Catalysis in the Decomposition of Diazo Compounds by Copper Chelates: Asymmetric Carbenoid Reactions," *Tetrahedron* 24:9 (1968), 3655–3669. | chiral,synthesis (45,954) |
| Physics 2005 | Contribution to the quantum theory of optical coherence | Glauber, Roy J., "The Quantum Theory of Optical Coherence," *Physical Review* 130:6 (1963), 2529. | optical_coherence (2,982) |
| Physics 2007 | Discovery of Giant Magnetoresistance | Baibich, Mário N., Jacques M. Broto, Albert Fert, Francis N. Van Dau, Frédéric Petroff, Paul Etienne, and J. Chazelas, "Giant Magnetoresistance of (001) Fe/(001) Cr Magnetic Superlattices," *Physical Review Letters* 61:21 (1988), 2472. | giant_magnetoresistance (3,229) |
| Medicine 2007 | Discoveries of principles for introducing specific gene modifications in mice by the use of embryonic stem cells | Thomas, Kirk R., and Mario R. Capecchi, "Site-Directed Mutagenesis by Gene Targeting in Mouse Embryo-Derived Stem Cells," *Cell* 51:3 (1987), 503–512. | gene_target (4,543) |
| Medicine 2009 | Discovery of how chromosomes are protected by telomeres and the enzyme telomerase | Szostak, Jack W., and Elizabeth H. Blackburn, "Cloning Yeast Telomeres on Linear Plasmid Vectors," *Cell* 29:1 (1982), 245–255. | polymerase,telomere (1,753) |
| Medicine 2011 | Discovery of the dendritic cell and its role in adaptive immunity | Steinman, Ralph M., and Zanvil A. Cohn, "Identification of a Novel Cell Type in Peripheral Lymphoid Organs of Mice. I. Morphology, Quantitation, Tissue Distribution," *The Journal of Experimental Medicine* 137:5 (1973), 1142–1162. | macrophage,dendritic_cell (8,299) |

*Notes:* 20 illustrative examples of Nobel prizes, a corresponding paper, and a new word (phrase) or new combination of words (phrases) introduced by the paper and found in the corresponding Nobel prize motivation page. Words (lemmatized) in phrases are separated by the underscore (e.g. 'optical_coherence' is a phrase) and new word (phrase) combinations are separated by the comma (e.g.'chiral,synthesis' is a new word combination). The reuse by later papers is shown in parentheses.

## Table 3: Predicting Nobel Prize Papers

**Panel A: Ex ante**

| | Text Metrics | | | | | Traditional Metrics | | Text Metrics Combined | All Metrics Combined |
|---|---|---|---|---|---|---|---|---|---|
| | (1) | (2) | (3) | (4) | (5) | (6) | (7) | (8) | (9) |
| New Word | 1.336*** | | | | | | | 0.243 | 0.257 |
| | (0.292) | | | | | | | (0.324) | (0.324) |
| New Phrase | | 1.906*** | | | | | | 1.498*** | 1.493*** |
| | | (0.193) | | | | | | (0.213) | (0.214) |
| New Word Combination | | | 0.675*** | | | | | 0.275** | 0.288** |
| | | | (0.090) | | | | | (0.112) | (0.112) |
| New Phrase Combination | | | | 0.984*** | | | | 0.596*** | 0.605*** |
| | | | | (0.139) | | | | (0.153) | (0.154) |
| Semantic Distance | | | | | -0.696 | | | -1.069 | -1.145 |
| | | | | | (1.352) | | | (1.460) | (1.459) |
| Wang | | | | | | 0.226 | | | -0.030 |
| | | | | | | (0.656) | | | (0.677) |
| Uzzi | | | | | | | -0.217 | | -0.369** |
| | | | | | | | (0.177) | | (0.182) |
| Log-Likelihood | -709.1 | -665.6 | -694.2 | -695.0 | -721.8 | -721.9 | -721.3 | -649.8 | -648.3 |
| Pseudo-$R^2$ | 0.113 | 0.167 | 0.131 | 0.130 | 0.097 | 0.097 | 0.098 | 0.187 | 0.189 |
| Precision (%) | 67.32 | 69.87 | 68.88 | 68.61 | 66.31 | 66.55 | 66.06 | 72.20 | 72.17 |
| Recall (%) | 65.45 | 65.63 | 64.93 | 65.28 | 63.89 | 63.54 | 63.19 | 67.19 | 67.53 |
| AUC | 0.7198 | 0.7635 | 0.7361 | 0.7362 | 0.7086 | 0.7083 | 0.7084 | 0.7745 | 0.7761 |
| Marginal Effects (%) | 9.09 | 20.93 | 23.05 | 32.15 | -0.81 | 0.64 | -1.78 | | |

**Panel B: Ex post**

| | Text Metrics | | | | Traditional Metrics | Text Metrics Combined | All Metrics Combined |
|---|---|---|---|---|---|---|---|
| | (10) | (11) | (12) | (13) | (14) | (15) | (16) |
| New Word Reuse | 0.338*** | | | | | 0.111* | 0.108* |
| | (0.052) | | | | | (0.065) | (0.064) |
| New Phrase Reuse | | 0.582*** | | | | 0.482*** | 0.471*** |
| | | (0.050) | | | | (0.055) | (0.055) |
| New Word Combination Reuse | | | 0.314*** | | | 0.169*** | 0.179*** |
| | | | (0.034) | | | (0.039) | (0.039) |
| New Phrase Combination Reuse | | | | 0.386*** | | 0.215*** | 0.199*** |
| | | | | (0.051) | | (0.058) | (0.058) |
| CD | | | | | 2.060*** | | 1.723*** |
| | | | | | (0.451) | | (0.473) |
| Log-Likelihood | -696.6 | -625.6 | -671.6 | -688.0 | -708.9 | -598.3 | -591.3 |
| Pseudo-$R^2$ | 0.128 | 0.217 | 0.160 | 0.139 | 0.113 | 0.251 | 0.260 |
| Precision (%) | 68.32 | 74.42 | 69.80 | 70.20 | 66.67 | 75.82 | 76.92 |
| Recall (%) | 64.41 | 67.19 | 67.01 | 67.88 | 63.89 | 68.58 | 69.44 |
| AUC | 0.7296 | 0.7938 | 0.7575 | 0.7470 | 0.7208 | 0.8139 | 0.8189 |
| Marginal Effects (%) | 13.23 | 25.66 | 23.02 | 23.31 | 7.85 | | |

*Notes:* Logit regressions with a binary outcome for Nobel Prize paper, robust standard errors in parentheses. Sample includes 1,168 papers (584 Nobel Prize papers matched with 584 control papers from the same year, journal, and subfield). All measures, except for *Semantic Distance* and *CD*, are log-transformed (after adding 1 for zero values). Models control for publication year and subfield fixed effects, abstract availability, text length (unique words and phrases in title and abstract), and number of unique papers and journals cited. AUC represents the area under the ROC curve. Marginal effects show the percentage increase in the likelihood of being a Nobel Prize paper associated with a one-standard-deviation increase in the metric. Baseline model (controls only): Precision = 66.25, Recall = 64.06, AUC = 0.7083. *** $p < 0.01$, ** $p < 0.05$, * $p < 0.10$.

# Online Appendix

**Appendix A**

**Figure A.1: Scientific Novelty by Selected Subfields of Physics and Astronomy and Year**

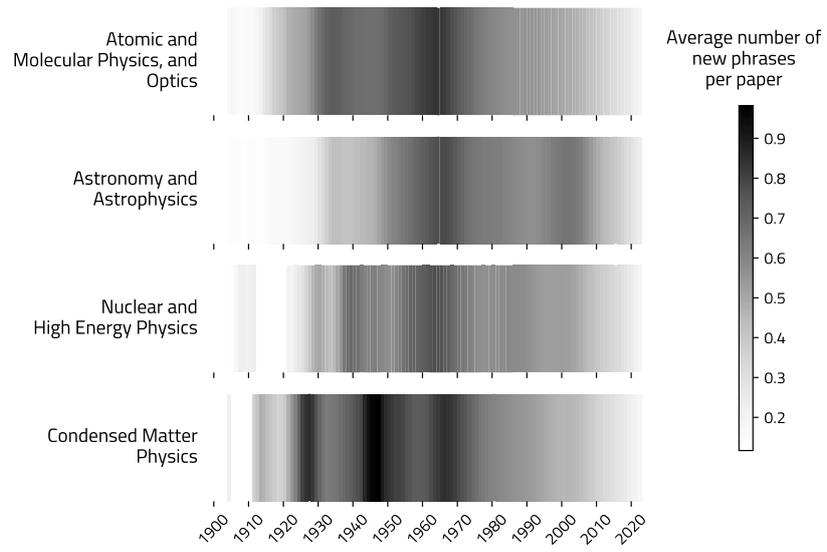

*Notes*: Paper-level average number of new phrases for a subset of 4 subfields of Physics and Astronomy (n=2,986,445 papers published between 1901 and 2023). Subfields are top down ordered by the overall number of publications.

**Table A.1: Summary Statistics of the Reuse of New Words, New Phrases and New Combinations of Words or Phrases**

| | # | Reuse | | | | | | | | | |
|---|---|---|---|---|---|---|---|---|---|---|---|
| | | Mean | Stdev | Min | p25 | p50 | p75 | p95 | p99 | Max | Skew |
| New Word | 6,285,104 | 39 | 1,239 | 1 | 1 | 2 | 7 | 57 | 400 | 633,484 | 223.691 |
| New Phrase | 27,079,343 | 17 | 435 | 1 | 1 | 2 | 5 | 34 | 194 | 416,734 | 315.428 |
| New Word Combination | 995,598,017 | 14 | 221 | 1 | 1 | 2 | 5 | 33 | 174 | 674,072 | 274.999 |
| New Phrase Combination | 824,362,160 | 5 | 47 | 1 | 1 | 1 | 3 | 15 | 58 | 128,896 | 294.534 |

*Notes:* Summary statistics of the reuse of new words, new phrases, and new combinations of words or phrases introduced from 1901 to 2023. # is the number of units (e.g. there are 6,285,104 new words introduced for the first time from 1901 to 2023 that are reused at least once). p25, p50, p75, p95 and p99 are respectively the $25^{th}$, $50^{th}$, $75^{th}$, $95^{th}$ and the $99^{th}$ percentile. The skewness (*skew*) of the distributions is the measure of the asymmetry of the probability distribution of a real-valued random variable about its mean. A positive skewness indicates that the distribution has a long right tail, while a negative skewness indicates a long left tail.

**Table A.2: Top 10 Reused New Noun Phrases by Decade**

| 1900-1910 | 1910-1920 | 1920-1930 | 1930-1940 | 1940-1950 | 1950-1960 |
|---|---|---|---|---|---|
| chemotherapy (334,823) | x-ray_diffraction (344,357) | antioxidant (273,607) | gene_expression (326,119) | genome (208,630) | transmission_electron_microscopy (188,353) |
| cardiovascular_disease (225,302) | peptide (271,529) | escherichia_coli (225,418) | ligand (303,778) | cytotoxicity (172,995) | neural_network (178,402) |
| antigen (220,632) | mitochondria (114,742) | myocardial_infarction (207,601) | thermal_stability (141,349) | mass_spectrometry (145,429) | liquid_chromatography (149,129) |
| radiotherapy (156,697) | atherosclerosis (107,828) | activation_energy (112,894) | lung_cancer (134,279) | prostate_cancer (131,725) | solar_cell (116,905) |
| schizophrenia (131,543) | squamous_cell_carcinoma (96,020) | monomer (110,685) | hydrogen_bond (126,377) | bandwidth (102,225) | immunohistochemical (100,576) |
| angiogenesis (99,680) | neutrophil (88,536) | temperature_dependence (109,640) | tomography (125,940) | spatial_resolution (97,117) | scan_electron_microscope (92,828) |
| inactivation (96,256) | surface_roughness (80,523) | electronic_structure (109,628) | electron_microscopy (110,050) | flavonoid (91,794) | monte_carlo_simulation (90,772) |
| cytotoxic (93,735) | isotope (66,655) | coronary_artery_disease (99,147) | chi-square (105,102) | surfactant (85,663) | anti-inflammatory (90,250) |
| photoluminescence (92,916) | energy_density (63,139) | atrial_fibrillation (95,099) | infrare_spectroscopy (104,334) | hepatitis_virus (81,529) | photocatalytic (87,497) |
| energy_efficiency (76,034) | stainless_steel (61,537) | blood_glucose (91,196) | insulin_resistance (100,685) | t-test (73,626) | cytometry (77,095) |

| 1960-1970 | 1970-1980 | 1980-1990 | 1990-2000 | 2000-2010 | 2010-2020 |
|---|---|---|---|---|---|
| scan_electron_microscopy (314,833) | apoptosis (416,734) | magnetic_resonance_image (233,975) | carbon_nanotube (74,429) | microrna (83,237) | blockchain (19,194) |
| mrna (308,514) | biomarker (325,535) | upregulation (120,043) | single_nucleotide_polymorphism (64,260) | mirnas (55,085) | covid-19_pandemic (12,904) |
| oxidative_stress (248,326) | nanoparticle (308,343) | il-6 (118,737) | genome-wide (61,234) | graphene_oxide (54,844) | abbvie (12,006) |
| immunohistochemistry (207,553) | body_mass_index (238,038) | polymerase_chain_reaction (116,146) | transcriptome (57,859) | metabolomic (40,035) | perovskite_solar_cell (10,440) |
| cytokine (205,311) | monoclonal_antibody (150,352) | comorbiditie (100,733) | nanomaterial (55,841) | kgaa (37,980) | circrna (9,474) |
| overexpression (186,933) | hplc (143,000) | nanocomposite (100,301) | convolutional_neural_network (54,526) | ma-seq (29,916) | deep_convolutional_neural_network (9,375) |
| oxygen_specie (170,168) | compute_tomography (129,159) | bioinformatic (90,771) | qrt-pcr (50,143) | lncrna (29,076) | radiomic (8,986) |
| transcription_factor (140,401) | flow_cytometry (118,743) | atomic_force_microscopy (85,709) | chemokine (48,459) | deep_neural_network (28,325) | ferroptosis (8,907) |
| colorectal_cancer (126,689) | mrna_expression (105,293) | nsclc (85,310) | fmri (43,226) | nafld (28,049) | generative_adversarial_network (8,832) |
| machine_learn (126,571) | molecular_dynamic_simulation (98,736) | egfr (81,671) | optical_coherence_tomography (42,324) | cloud_compute (25,861) | mxene (8,607) |

*Notes:* Top 10 reused new noun phrases by decade. Lemmatized words in phrases are separated by the underscore. The reuse by later papers is shown in parentheses.

**Table A.3: Correlation matrix**

| | (1) | (2) | (3) | (4) | (5) | (6) | (7) | (8) | (9) | (10) | (11) | (12) | (13) | (14) | (15) | (16) | (17) | (18) | (19) | (20) |
|---|---|---|---|---|---|---|---|---|---|---|---|---|---|---|---|---|---|---|---|---|
| (1) New Word (Binary) | 1.00 | | | | | | | | | | | | | | | | | | | |
| (2) New Word | 0.86 | 1.00 | | | | | | | | | | | | | | | | | | |
| (3) New Phrase (Binary) | 0.31 | 0.29 | 1.00 | | | | | | | | | | | | | | | | | |
| (4) New Phrase | 0.36 | 0.41 | 0.81 | 1.00 | | | | | | | | | | | | | | | | |
| (5) New Word Combination (Binary) | 0.20 | 0.17 | 0.33 | 0.29 | 1.00 | | | | | | | | | | | | | | | |
| (6) New Word Combination | 0.14 | 0.27 | 0.13 | 0.28 | 0.11 | 1.00 | | | | | | | | | | | | | | |
| (7) New Phrase Combination (Binary) | 0.14 | 0.12 | 0.31 | 0.27 | 0.60 | 0.09 | 1.00 | | | | | | | | | | | | | |
| (8) New Phrase Combination | 0.13 | 0.20 | 0.19 | 0.35 | 0.17 | 0.67 | 0.18 | 1.00 | | | | | | | | | | | | |
| (9) Semantic Distance | 0.03 | 0.03 | 0.07 | 0.06 | 0.13 | 0.01 | 0.09 | -0.02 | 1.00 | | | | | | | | | | | |
| (10) Wang | 0.00 | 0.00 | 0.01 | 0.01 | 0.02 | 0.01 | 0.03 | 0.01 | -0.01 | 1.00 | | | | | | | | | | |
| (11) Uzzi | 0.00 | 0.00 | 0.01 | 0.01 | 0.01 | 0.00 | 0.01 | 0.00 | -0.01 | -0.01 | 1.00 | | | | | | | | | |
| (12) New Word Reuse | 0.03 | 0.04 | 0.01 | 0.02 | 0.01 | 0.01 | 0.00 | 0.01 | 0.00 | 0.00 | 0.00 | 1.00 | | | | | | | | |
| (13) New Phrase Reuse | 0.03 | 0.03 | 0.04 | 0.06 | 0.01 | 0.01 | 0.01 | 0.02 | 0.00 | 0.00 | 0.00 | 0.44 | 1.00 | | | | | | | |
| (14) New Word Combination Reuse | 0.09 | 0.13 | 0.10 | 0.19 | 0.07 | 0.24 | 0.06 | 0.24 | 0.00 | 0.00 | 0.00 | 0.23 | 0.22 | 1.00 | | | | | | |
| (15) New Phrase Combination Reuse | 0.11 | 0.15 | 0.16 | 0.27 | 0.12 | 0.36 | 0.12 | 0.60 | -0.03 | 0.01 | -0.01 | 0.07 | 0.15 | 0.56 | 1.00 | | | | | |
| (16) CD | 0.01 | 0.01 | 0.01 | 0.02 | 0.00 | 0.00 | 0.00 | 0.00 | 0.01 | -0.01 | 0.00 | 0.00 | 0.01 | 0.02 | 0.01 | 1.00 | | | | |
| (17) Abstract | 0.09 | 0.08 | 0.19 | 0.18 | 0.44 | 0.07 | 0.50 | 0.12 | 0.17 | 0.02 | 0.01 | 0.00 | 0.01 | 0.04 | 0.08 | 0.00 | 1.00 | | | |
| (18) N. of Words | 0.15 | 0.17 | 0.26 | 0.31 | 0.48 | 0.34 | 0.53 | 0.44 | 0.10 | 0.03 | -0.02 | 0.00 | 0.01 | 0.11 | 0.24 | -0.02 | 0.49 | 1.00 | | |
| (19) N. of Phrases | 0.18 | 0.20 | 0.30 | 0.36 | 0.50 | 0.34 | 0.55 | 0.51 | 0.04 | 0.02 | -0.02 | 0.00 | 0.01 | 0.12 | 0.28 | -0.02 | 0.45 | 0.92 | 1.00 | |
| (20) N. of cited Papers | -0.02 | -0.02 | 0.01 | 0.01 | 0.11 | 0.01 | 0.17 | 0.04 | 0.05 | 0.21 | -0.04 | 0.00 | -0.01 | -0.01 | 0.01 | -0.05 | 0.18 | 0.27 | 0.28 | 1.00 |
| (21) N. of cited Journals | -0.03 | -0.03 | -0.01 | -0.01 | 0.12 | 0.00 | 0.18 | 0.04 | 0.08 | 0.20 | -0.06 | 0.00 | -0.01 | -0.02 | 0.00 | -0.05 | 0.20 | 0.31 | 0.31 | 0.90 |

*Notes:* n= 75,295,921 papers published between 1901 and 2023.

**Table A.4: Descriptive Statistics Nobel Prize versus Control Papers**

| | Mean | Stdev | Min | p25 | p50 | p75 | p95 | p99 | Max | Mean | Stdev | Min | p25 | p50 | p75 | p95 | p99 | Max | Z | p-value |
|---|---|---|---|---|---|---|---|---|---|---|---|---|---|---|---|---|---|---|---|---|
| | Nobel prize papers (n=584) | | | | | | | | | Control papers (n=584) | | | | | | | | | | |
| **Panel A: Ex ante** | | | | | | | | | | | | | | | | | | | | |
| New Word (Binary) | 0.223 | 0.416 | 0.000 | 0.000 | 0.000 | 0.000 | 1.000 | 1.000 | 1.000 | 0.084 | 0.277 | 0.000 | 0.000 | 0.000 | 0.000 | 1.000 | 1.000 | 1.000 | -6.577 | 0.0000*** |
| New Word | 0.187 | 0.380 | 0.000 | 0.000 | 0.000 | 0.000 | 1.099 | 1.609 | 2.398 | 0.066 | 0.226 | 0.000 | 0.000 | 0.000 | 0.000 | 0.693 | 1.099 | 1.609 | -6.601 | 0.0000*** |
| New Phrase (Binary) | 0.551 | 0.498 | 0.000 | 0.000 | 1.000 | 1.000 | 1.000 | 1.000 | 1.000 | 0.236 | 0.425 | 0.000 | 0.000 | 0.000 | 0.000 | 1.000 | 1.000 | 1.000 | -11.014 | 0.0000*** |
| New Phrase | 0.590 | 0.631 | 0.000 | 0.000 | 0.693 | 1.099 | 1.792 | 2.303 | 2.890 | 0.199 | 0.380 | 0.000 | 0.000 | 0.000 | 0.000 | 1.099 | 1.386 | 1.609 | -11.881 | 0.0000*** |
| New Word Combination (Binary) | 0.671 | 0.470 | 0.000 | 0.000 | 1.000 | 1.000 | 1.000 | 1.000 | 1.000 | 0.452 | 0.498 | 0.000 | 0.000 | 0.000 | 1.000 | 1.000 | 1.000 | 1.000 | -7.545 | 0.0000*** |
| New Word Combination | 1.895 | 1.795 | 0.000 | 0.000 | 1.386 | 3.497 | 4.913 | 5.631 | 7.258 | 0.927 | 1.302 | 0.000 | 0.000 | 0.000 | 1.609 | 3.761 | 4.575 | 5.814 | -9.636 | 0.0000*** |
| New Phrase Combination (Binary) | 0.738 | 0.440 | 0.000 | 0.000 | 1.000 | 1.000 | 1.000 | 1.000 | 1.000 | 0.509 | 0.500 | 0.000 | 0.000 | 1.000 | 1.000 | 1.000 | 1.000 | 1.000 | -8.088 | 0.0000*** |
| New Phrase Combination | 1.873 | 1.692 | 0.000 | 0.000 | 1.386 | 3.296 | 4.812 | 5.447 | 6.534 | 0.967 | 1.279 | 0.000 | 0.000 | 0.693 | 1.609 | 3.611 | 4.787 | 5.684 | -9.773 | 0.0000*** |
| Semantic Distance | 0.137 | 0.054 | 0.010 | 0.102 | 0.131 | 0.160 | 0.218 | 0.293 | 0.677 | 0.137 | 0.053 | 0.010 | 0.101 | 0.130 | 0.166 | 0.229 | 0.298 | 0.425 | 0.109 | 0.9128 |
| Wang | 0.029 | 0.150 | 0.000 | 0.000 | 0.000 | 0.000 | 0.000 | 0.877 | 1.389 | 0.011 | 0.100 | 0.000 | 0.000 | 0.000 | 0.000 | 0.000 | 0.487 | 1.487 | -2.836 | 0.0046*** |
| Uzzi | 4.068 | 0.376 | -0.018 | 4.043 | 4.043 | 4.043 | 4.513 | 5.864 | 6.772 | 4.094 | 0.368 | 1.209 | 4.043 | 4.043 | 4.043 | 4.689 | 6.079 | 6.713 | 0.615 | 0.5384 |
| **Panel B: Ex post** | | | | | | | | | | | | | | | | | | | | |
| New Word Reuse | 1.100 | 2.429 | 0.000 | 0.000 | 0.000 | 0.000 | 7.088 | 9.847 | 11.806 | 0.251 | 0.952 | 0.000 | 0.000 | 0.000 | 0.000 | 2.398 | 5.278 | 8.097 | -6.931 | 0.0000*** |
| New Phrase Reuse | 2.531 | 2.836 | 0.000 | 0.000 | 1.792 | 4.543 | 8.047 | 10.013 | 11.663 | 0.571 | 1.235 | 0.000 | 0.000 | 0.000 | 0.000 | 3.367 | 5.357 | 7.754 | -13.198 | 0.0000*** |
| New Word Combination Reuse | 4.801 | 3.940 | 0.000 | 0.000 | 5.407 | 8.077 | 10.574 | 12.230 | 13.251 | 2.264 | 2.898 | 0.000 | 0.000 | 0.000 | 4.718 | 7.826 | 9.534 | 11.047 | -11.238 | 0.0000*** |
| New Phrase Combination Reuse | 3.886 | 3.078 | 0.000 | 0.000 | 3.989 | 6.531 | 8.564 | 9.750 | 10.897 | 2.050 | 2.475 | 0.000 | 0.000 | 1.099 | 3.922 | 6.632 | 8.322 | 10.919 | -10.523 | 0.0000*** |
| CD | 0.092 | 0.203 | -0.461 | 0.000 | 0.000 | 0.084 | 0.595 | 0.901 | 0.968 | 0.047 | 0.144 | -0.200 | 0.000 | 0.000 | 0.006 | 0.400 | 0.759 | 0.982 | -4.559 | 0.0000*** |

*Notes:* n=1,168 papers of which 584 Nobel prize papers and 584 matched control papers. Each Nobel prize paper is matched to one randomly selected control paper published in the same year, journal and subfield. p25, p50, p75, p95 and p99 are respectively the 25[th], 50[th], 75[th], 95[th] and the 99[th] percentile. All measures except *Semantic Distance* and *CD* are log transformed after adding 1 for measures with 0 values. Z values are test statistics from the Mann-Whitney test. *** p<0.01, ** p<0.05, * p<0.10.

**Appendix B. Text Metrics Exclusively based on Titles**

Abstracts are unavailable for a significant portion of papers in OpenAlex, which may introduce bias in the text metrics and analyses. To address this, we include a binary indicator for abstract availability in all regressions and control for the text length of each paper (measured by the number of words and phrases), as longer texts may increase the likelihood of introducing new words, phrases, or their combinations. As an additional robustness check, we recalculated all text metrics (excluding the embedding-based metric, *Semantic Distance*) for the full OpenAlex dataset using only paper titles. We then replicated the Nobel Prize validation using these title-based metrics.[10] Tables B.1 and B.2 present the descriptive statistics and regression results. Our main findings remain consistent, with all text-based metrics outperforming traditional citation-based metrics. However, as expected, the overall predictive power of title-based metrics is lower, underscoring the additional value that paper abstracts provide in identifying new scientific ideas and measuring their impact on subsequent work.

---

[10] We find similar results in the validation based on literature review papers. These results are not shown here but are available upon request from the authors.

**Table B.1: Descriptive Statistics Nobel Prize versus Control Papers (only titles)**

| | Mean | Stdev | Min | p25 | p50 | p75 | p95 | p99 | Max | Mean | Stdev | Min | p25 | p50 | p75 | p95 | p99 | Max | Z | p-value |
|---|---|---|---|---|---|---|---|---|---|---|---|---|---|---|---|---|---|---|---|---|
| | Nobel prize papers (n=584) | | | | | | | | | Control papers (n=584) | | | | | | | | | | |
| **Panel A: Ex ante** | | | | | | | | | | | | | | | | | | | | |
| New Word (Binary) | 0.094 | 0.292 | 0.000 | 0.000 | 0.000 | 0.000 | 1.000 | 1.000 | 1.000 | 0.053 | 0.224 | 0.000 | 0.000 | 0.000 | 0.000 | 1.000 | 1.000 | 1.000 | -2.688 | 0.0072*** |
| New Word | 0.070 | 0.222 | 0.000 | 0.000 | 0.000 | 0.000 | 0.693 | 1.099 | 1.386 | 0.039 | 0.167 | 0.000 | 0.000 | 0.000 | 0.000 | 0.693 | 0.693 | 1.099 | -2.692 | 0.0071*** |
| New Phrase (Binary) | 0.322 | 0.468 | 0.000 | 0.000 | 0.000 | 1.000 | 1.000 | 1.000 | 1.000 | 0.187 | 0.390 | 0.000 | 0.000 | 0.000 | 0.000 | 1.000 | 1.000 | 1.000 | -5.306 | 0.0000*** |
| New Phrase | 0.238 | 0.356 | 0.000 | 0.000 | 0.000 | 0.693 | 0.693 | 1.099 | 1.609 | 0.133 | 0.282 | 0.000 | 0.000 | 0.000 | 0.000 | 0.693 | 0.693 | 1.386 | -5.430 | 0.0000*** |
| New Word Combination (Binary) | 0.521 | 0.500 | 0.000 | 0.000 | 1.000 | 1.000 | 1.000 | 1.000 | 1.000 | 0.389 | 0.488 | 0.000 | 0.000 | 0.000 | 1.000 | 1.000 | 1.000 | 1.000 | -4.523 | 0.0000*** |
| New Word Combination | 0.610 | 0.691 | 0.000 | 0.000 | 0.693 | 1.099 | 1.946 | 2.398 | 3.526 | 0.432 | 0.623 | 0.000 | 0.000 | 0.000 | 0.693 | 1.792 | 2.398 | 3.178 | -4.755 | 0.0000*** |
| New Phrase Combination (Binary) | 0.401 | 0.490 | 0.000 | 0.000 | 0.000 | 1.000 | 1.000 | 1.000 | 1.000 | 0.223 | 0.416 | 0.000 | 0.000 | 0.000 | 0.000 | 1.000 | 1.000 | 1.000 | -6.567 | 0.0000*** |
| New Phrase Combination | 0.388 | 0.539 | 0.000 | 0.000 | 0.000 | 0.693 | 1.386 | 1.946 | 2.944 | 0.188 | 0.377 | 0.000 | 0.000 | 0.000 | 0.000 | 1.099 | 1.386 | 1.946 | -6.985 | 0.0000*** |
| Wang | 0.029 | 0.150 | 0.000 | 0.000 | 0.000 | 0.000 | 0.000 | 0.877 | 1.389 | 0.011 | 0.100 | 0.000 | 0.000 | 0.000 | 0.000 | 0.000 | 0.487 | 1.487 | -2.836 | 0.0046*** |
| Uzzi | 4.068 | 0.376 | -0.018 | 4.043 | 4.043 | 4.043 | 4.513 | 5.864 | 6.772 | 4.094 | 0.368 | 1.209 | 4.043 | 4.043 | 4.043 | 4.689 | 6.079 | 6.713 | 0.615 | 0.5384 |
| **Panel B: Ex post** | | | | | | | | | | | | | | | | | | | | |
| New Word Reuse | 0.446 | 1.598 | 0.000 | 0.000 | 0.000 | 0.000 | 3.638 | 8.186 | 10.999 | 0.113 | 0.544 | 0.000 | 0.000 | 0.000 | 0.000 | 1.099 | 2.773 | 5.866 | -2.896 | 0.0038*** |
| New Phrase Reuse | 1.180 | 2.085 | 0.000 | 0.000 | 0.000 | 1.792 | 5.849 | 8.558 | 10.643 | 0.467 | 1.181 | 0.000 | 0.000 | 0.000 | 0.000 | 3.367 | 5.903 | 7.719 | -6.079 | 0.0000*** |
| New Word Combination Reuse | 2.330 | 2.664 | 0.000 | 0.000 | 1.242 | 4.615 | 6.972 | 9.088 | 10.353 | 1.294 | 1.971 | 0.000 | 0.000 | 0.000 | 2.350 | 5.635 | 7.754 | 8.377 | -6.464 | 0.0000*** |
| New Phrase Combination Reuse | 0.940 | 1.378 | 0.000 | 0.000 | 0.000 | 1.609 | 3.871 | 5.509 | 6.706 | 0.430 | 0.905 | 0.000 | 0.000 | 0.000 | 0.000 | 2.398 | 3.912 | 5.220 | -7.041 | 0.0000*** |
| CD | 0.092 | 0.203 | -0.461 | 0.000 | 0.000 | 0.084 | 0.595 | 0.901 | 0.968 | 0.047 | 0.144 | -0.200 | 0.000 | 0.000 | 0.006 | 0.400 | 0.759 | 0.982 | -4.559 | 0.0000*** |

*Notes*: n=1,168 papers of which 584 Nobel prize papers and 584 matched control papers. Each Nobel prize paper is matched to one randomly selected control paper published in the same year, journal and subfield. p25, p50, p75, p95 and p99 are respectively the $25^{th}$, $50^{th}$, $75^{th}$, $95^{th}$ and the $99^{th}$ percentile. All measures except binary indicators and CD are log transformed after adding 1 for measures with 0 values. Z values are test statistics from the Mann-Whitney test. *** p<0.01, ** p<0.05, * p<0.10.

## Table B.2: Predicting Nobel Prize Papers (only titles)

| | Text Metrics | | | | Traditional Metrics | | Text Metrics Combined | All Metrics Combined |
|---|---|---|---|---|---|---|---|---|
| **Panel A: Ex ante** | | | | | | | | |
| | (1) | (2) | (3) | (4) | (5) | (6) | (7) | (8) |
| New Word | 1.198*** | | | | | | 0.702 | 0.730* |
| | (0.399) | | | | | | (0.436) | (0.434) |
| New Phrase | | 1.210*** | | | | | 0.929*** | 0.925*** |
| | | (0.230) | | | | | (0.250) | (0.252) |
| New Word Combination | | | 0.456*** | | | | 0.108 | 0.106 |
| | | | (0.132) | | | | (0.143) | (0.143) |
| New Phrase Combination | | | | 1.129*** | | | 1.018*** | 1.020*** |
| | | | | (0.177) | | | (0.183) | (0.184) |
| Wang | | | | | 0.448 | | | 0.490 |
| | | | | | (0.593) | | | (0.689) |
| Uzzi | | | | | | -0.149 | | -0.136 |
| | | | | | | (0.182) | | (0.187) |
| Log-Likelihood | -731.4 | -721.7 | -730.5 | -736.5 | -736.5 | -721.5 | -702.4 | -701.7 |
| Pseudo-$R^2$ | 0.085 | 0.097 | 0.086 | 0.104 | 0.078 | 0.097 | 0.121 | 0.122 |
| Precision (%) | 62.46 | 64.24 | 63.41 | 66.03 | 62.30 | 62.26 | 67.67 | 68.32 |
| Recall (%) | 62.67 | 64.24 | 63.19 | 66.15 | 61.98 | 62.15 | 66.49 | 67.01 |
| AUC | 0.6908 | 0.7053 | 0.6901 | 0.7114 | 0.6801 | 0.6796 | 0.7302 | 0.7315 |
| Marginal Effects (%) | 5.19 | 8.56 | 6.71 | 11.60 | 1.29 | -1.25 | | |
| **Panel B: Ex post** | | | | | | | | |
| | (9) | (10) | (11) | (12) | (13) | | (14) | (15) |
| New Word Reuse | 0.392*** | | | | | | 0.280*** | 0.252*** |
| | (0.078) | | | | | | (0.091) | (0.088) |
| New Phrase Reuse | | 0.332*** | | | | | 0.209*** | 0.210*** |
| | | (0.046) | | | | | (0.051) | (0.051) |
| New Word Combination Reuse | | | 0.252*** | | | | 0.170*** | 0.166*** |
| | | | (0.036) | | | | (0.038) | (0.039) |
| New Phrase Combination Reuse | | | | 0.476*** | | | 0.366*** | 0.354*** |
| | | | | (0.070) | | | (0.070) | (0.072) |
| CD | | | | | 2.012*** | | | 1.822*** |
| | | | | | (0.448) | | | (0.453) |
| Log-Likelihood | -721.5 | -707.8 | -710.1 | -712.7 | -724.2 | | -675.2 | -665.9 |
| Pseudo-$R^2$ | 0.097 | 0.114 | 0.112 | 0.108 | 0.094 | | 0.155 | 0.167 |
| Precision (%) | 63.46 | 64.85 | 65.68 | 65.84 | 62.76 | | 69.31 | 69.86 |
| Recall (%) | 61.81 | 64.06 | 65.45 | 64.58 | 63.19 | | 66.67 | 67.19 |
| AUC | 0.7001 | 0.7174 | 0.7180 | 0.7132 | 0.6953 | | 0.7539 | 0.7614 |
| Marginal Effects (%) | 9.99 | 12.20 | 12.87 | 12.17 | 7.87 | | | |

*Notes:* Logit regressions with a binary outcome for Nobel Prize paper, robust standard errors in parentheses. Sample includes 1,168 papers (584 Nobel Prize papers matched with 584 control papers from the same year, journal, and subfield). All measures, except for *CD*, are log-transformed (after adding 1 for zero values). Models control for publication year and subfield fixed effects, abstract availability, text length (unique words and phrases in title), and number of unique papers and journals cited. AUC represents the area under the ROC curve. Marginal effects show the percentage increase in the likelihood of being a Nobel Prize paper associated with a one-standard-deviation increase in the metric. Baseline model (controls only): Precision=60.78, recall=62.15, AUC=0.6694. *** p<0.01, ** p<0.05, * p<0.10.

**Appendix C. Data Collection and Processing**

We collect publications from the OpenAlex snapshot, dated 2024.01.09[11] (Priem, Piwowar, & Orr, 2022). OpenAlex contains 246,880,876 records including books, book chapters, conference proceedings, datasets, journal articles, papers, repositories, theses and preprints. We download the sample of English and non-retracted journal publications and published conference proceedings (112,118,702 records),[12] referred to as "publications" in the remainder of the paper.

      Next, we remove 9,118,483 publications that have a non-English title or abstract,[13] or an abstract that includes both English and a translated version, 20,208,011 publications published in journals or conference proceedings that have no publisher or impact factor, 7,853,328 publications with duplicate titles, 86,056 publications with an empty title and 5,702,538 publications with no authors. Moreover, we only keep the title and drop the abstract for 5,436,159 publications that have duplicated abstracts. We also drop the abstract for 7,511,164 publications that mainly contain bibliographic data (i.e. references) in the abstract. Finally, we remove 3,054 publications that were misclassified as introducing new words due to incorrect data (e.g., non-existing URLs, wrong dates, or wrong titles),[14] and we exclude the abstract for 8,420 papers where OpenAlex contains inaccuracies, such as abstracts that do not match the actual publication[15]. We ensure the validity of these restrictions through random manual checks at each step, documenting the process in greater detail in the documentation in our online data repository (https://zenodo.org/records/13869486). The final sample consists of 75,612,702 publications from 1666 to 2023.

      Next, we process the title and abstracts of these publications. For every publication, the abstract is only available as an inverted index in OpenAlex, which we convert back to the original abstract.[16] Because there is only a title and no abstract available for 20% of the publications, we illustrate in Appendix B the robustness of all metrics and findings in case we discard abstracts and exclusively rely on titles. Afterwards, we concatenate the title and abstract and process the text using the spacy model "*en_core_sci_lg*"[17] with the following steps: tokenization, POS-tagging, dependency parsing, chunking, lemmatization, cleaning, baseline removal, and vectorization. First, we lowercase the text and tokenize it to individual words (or unigrams).[18] Second, we apply Part-of-Speech (POS) tagging to classify each word based on its definition and sentence context.[19] Third, we use dependency parsing to determine the grammatical structure of each sentence and establish relationships between words, showing how they are grammatically connected. Fourth, we perform chunking by extracting noun phrases. Noun phrases consist of one or more words with a noun as their head, essentially a noun plus its descriptors. We identify noun phrases by examining the grammatical categories and dependencies recognized by the POS-tagging and the dependency parser. This involves locating nouns and proper nouns, then collecting all grammatically dependent words. Noun phrases can range

---

[11] https://openalex.s3.amazonaws.com/browse.html#data/works/

[12] Specifically, we applied the filter on the "source type" to be "journal" or "conference" and to "language" to be "en".

[13] OpenAlex assigns the language based on the words in the publication's abstract, or the title if the abstract is unavailable. This may result in cases where the title is in a different language from the abstract. To mitigate this issue, we separately assess the language of the title and abstract and exclude papers where either is not in English. We use the spacy FastLang detector (https://spacy.io/universe/project/spacy_fastlang).

[14] Another aspect to consider is that OpenAlex assigns the 1st of January as the publication date for 30,079,435 papers, a significantly large share of papers. However, 41% of these papers contain issues such as missing publishers, no impact-factor, non-English titles or duplicate titles. After excluding these papers with potential inconsistencies, we randomly checked 100 of the remaining papers dated 1st of January and we found that 85% has the correct publication date, while 15% lacks sufficient details to confirm the exact date (e.g. only the year is mentioned). Since potential errors are already addressed through prior exclusions and that OpenAlex already cross-references dates with other sources like CrossRef and PubMed, we decide to retain these papers dated the 1st of January.

[15] For these manual checks, we focused on the top reused 5,000 newly introduced words (after 1900) and checked the first five publications associated with each new word. We examined their titles, abstracts, publication dates, DOIs, and journal information, manually verifying whether the data were accurate or the papers were correctly identified as introducing these new words. Our assumption is that if a record contains incorrect data, it is highly likely to be misclassified as the first to introduce a new word. See the data repository documentation for more details about this process.

[16] Inverted indexes store information about each word in a body of text, i.e. they map the content of the body of text to their relative location. This includes the number of occurrences a word is found and its position of occurrence.

[17] This is a spacy model tailored for processing scientific text (Neumann et al., 2019)

[18] We use the scispacy tokenizer, in order to consider words containing hyphens ("-") as single unigrams.

[19] POS (Part-of-Speech) tagging classifies words into categories such as nouns, verbs, adjectives, etc.

from single words (with no dependencies) to full sentences.[20] Next, we apply lemmatization to each word using the scispacy lemmatizer so that we are left with a collection of lemmas representing the scientific content of the publication.[21] [22]

We then proceed with a cleaning step. First, we collect the 74,579 most common (lemmatized) words appearing in at least 1,000 papers.[23] From this list, we identify two categories of words: *stop words* (n=2,185) and *removal words* (n=7,309). Stop words lack scientific meaning, both individually and in combination with other words (e.g., "abstract," "university," "journal"). Removal words lack scientific meaning alone and when combined with each other, but gain scientific meaning when combined with other words (e.g., "autonomous," "conservative," "movement"). Next, we expand these lists. Hyphenated words containing at least one stop word are considered *stop words*, while hyphenated words made entirely of removal words are considered *removal words*. Additionally, we further expand the *stop words* list by adding natural stop words,[24] words composed only by numbers and potentially mistakenly combined words.[25] The final list of *stop words* consists of 1,542,813 words, while the final list of *removal words* consists of 326,503 words. Finally, for each publication, we remove noun phrases, word combinations and phrase combinations that contain any stop word or consist entirely of removal words. However, we retain noun phrases after removing any leading stop words, unless the remaining phrase consists solely of removal words. Finally, we remove phrase combinations where one phrase is an acronym formed from the first letters of the other phrase.

Finally, we create a vector representation of the scientific content (title and abstract not pre-processed) of every publication using SPECTER, a pre-trained document-level embedding of scientific documents that is pretrained on a Transformer language model tailored to scientific text. SPECTER is currently the best performing document-level embedding of scientific papers and outperforms other embedding models such as SciBERT on different benchmark tasks. For every publication, we create a vector of 768 dimensions using the SPECTER public API.[26]

We use the title, abstract, and full text of all publications from 1666 to 1900 to construct a baseline dictionary (n=316,781). Among these, the full text is available for 129,738 publications published before 1901.[27] Our analysis is restricted to papers published between 1901 and 2023 (n=75,295,921). The entire cleaned vocabulary contains 19,787,208 unique words (unigrams) and 104,613,632 unique noun phrases. The average and median number of unique words per publication is 32 and 30 and the average and median number of unique phrases per publication is 13 and 11.

---

[20] For instance, in the title text "Specific Enzymatic Amplification of DNA In Vitro: The Polymerase Chain Reaction" (Mullis et al. 1986), we identify the nouns "amplification", "dna", "vitro", "polymerase," "chain," and "reaction." POS-tagging and dependency parsing classifies "specific" and "enzymatic" as adjectives modifying "amplification," forming the noun phrase "specific enzymatic amplification", while "polymerase" and "chain" form a compound noun with "reaction", resulting in the noun phrase "polymerase chain reaction". Therefore, the noun phrases identified are "specific enzymatic amplification", "dna", "vitro" and "polymerase chain reaction".

[21] We also lemmatize the constituent words of hyphenated unigrams.

[22] When using POS-tagging and dependency parsing, words can be tagged or lemmatized differently based on their context (Neumann et al., 2019). For instance, past-participle verbs used as adjectives, gerundive verbs used as nouns, or polysemic words may be processed differently depending on their specific usage in the text (e.g., the word "aids" could refer to the verb "to aid" or the noun referring to the disease "AIDS"). These contextual variations affect the overall accuracy of the POS-tagging, dependency parsing and lemmatization, as well as the extraction of the noun phrases (F1 score = 68.97%) (Neumann et al., 2019). Although we use scispacy, that is the best performing state-of-the-art model for processing scientific text, there might be potential errors in text processing.

[23] 1,000 is a reasonable threshold since it captures words used in at least 0.0013% of the total sample.

[24] We consider natural stop words as: (i) numbers or words referring to numbers written in full (i.e. one, two, etc.); (ii) stop words collected from the NLTK Python library (127 words), stop words collected from Gensim Python library (337 words), stop words collected from Spacy Python library (326 words) (iv) words referring to days and months (i.e. Monday, July, etc.); and (vi) words referring to cities (i.e. Boston, London, etc.), countries (i.e. USA, Europe, etc.), first names (i.e. John, Marie, etc.) and institutions (i.e., Harvard, Stanford, etc.) retrieved from Wikidata (8,570 words).

[25] For example, words that start or end with an hyphen or contain double hyphens ("--"), non-ascii words etc. We expand this list also with the stop words provided by Arts, Hou, & Gomez (2021).

[26] https://github.com/allenai/paper-embedding-public-apis

[27] OpenAlex allows retrieving full text from the Internet Archive, but it can only be accessed via the API, not as a single snapshot. This requires retrieving each paper's text one-by-one, making it time-consuming and impractical for all papers. For this reason, we retrieve only the full text of publications before 1900. Additionally, the API provides the text in the form of n-grams instead of plain text, limiting our ability to fully process it. Therefore, we only apply lemmatization and cleaning to these papers.

**Appendix D. Citations to Papers Introducing New Scientific Ideas**

We randomly sampled 10,000 new words that were reused at least 10 times in subsequent papers, identifying 2,281,960 reusing papers. Each reusing paper was matched to a randomly selected control paper that did not reuse the specific word but was published in the same journal, year, and subfield. If no match was available at the subfield level, we matched based on journal, year, and field. Similarly, we sampled 10,000 new phrases reused at least 10 times, identifying 1,007,422 reusing papers; 10,000 new word combinations reused at least 10 times, identifying 1,049,192 reusing papers; and 10,000 new phrase combinations reused at least 10 times, identifying 410,406 reusing papers. For each case, reusing papers were matched to corresponding control papers. Papers reusing new words were 53 times more likely to cite the paper that first introduced the word compared to the matched control group (0.854% versus 0.016%). Similarly, papers reusing new phrases were 45 times more likely (0.939% versus 0.021%), papers reusing new word combinations were 40 times more likely (0.561% versus 0.014%), and papers reusing new phrase combinations were 21 times more likely (0.972% versus 0.047%). These findings suggest that the reuse of new ideas correlates with citations to the paper pioneering these ideas, confirming their intellectual influence. However, the data also reveal that only a small minority of reusing papers cite the pioneering paper, emphasizing the value of text-based metrics in providing a complementary and novel perspective on the diffusion and influence of new ideas.

To analyze how the likelihood of citation evolves over time, we estimate linear probability models for each metric, focusing on the probability that reusing papers (those reusing a new scientific idea) cite the pioneering papers (those introducing the idea). The dependent variable in each model is a binary indicator of whether a reusing paper cites the pioneering paper, while the key independent variable is the time difference (in years) between their publication dates, modeled as a series of year-specific indicators (with 0 years as the reference category). All models include fixed effects for the subfield of both the pioneering and reusing papers, as well as publication year fixed effects for the pioneering paper. Additional controls for both the pioneering and reusing papers include whether an abstract is available, text length (number of unique words and phrases in the title and abstract), and the number of unique papers and journals cited. Figure D.1 shows that the predicted probabilities of a reusing paper citing the corresponding pioneering paper decrease significantly over time.

**Figure D.1: Predicted Probability of Citation for Reusing Papers**

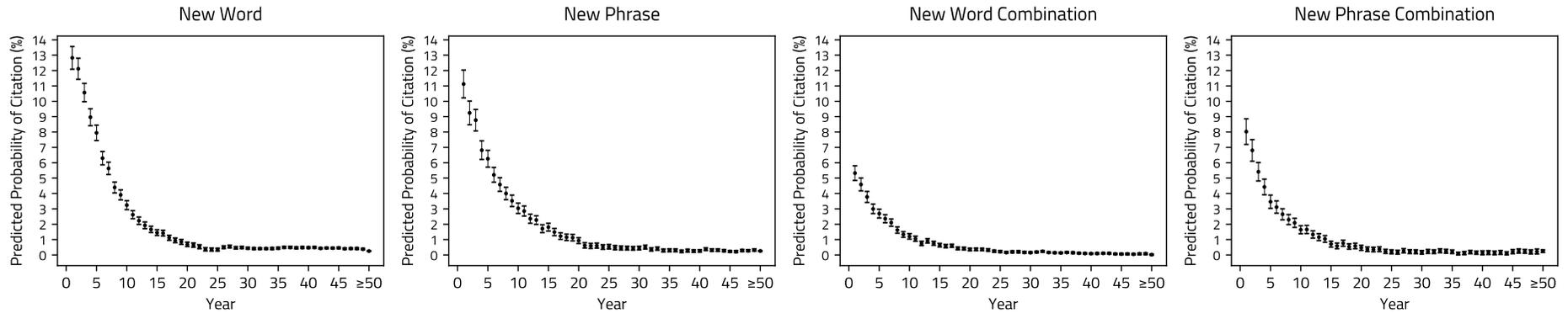

*Notes*: This figure plots the predicted probability (in %) that reusing papers (those reusing a new idea, such as a new word or phrase) cite their corresponding pioneering papers (those introducing the idea) over time, with 95% confidence intervals. Predictions are derived from separate linear probability models estimated for each metric, where the dependent variable indicates whether a reusing paper cites the pioneering paper. The key independent variable is the time difference (in years) between the publication dates of the pioneering and reusing papers, modeled as a series of year-specific indicators (0 years as the reference category). Predicted probabilities are calculated for each year difference. For *New Word*, the sample includes 2,281,960 reusing papers involving 10,000 randomly sampled new words reused at least 10 times. For *New Phrase*, the sample includes 1,007,422 reusing papers involving 10,000 randomly sampled new phrases reused at least 10 times. For *New Word Combination*, the sample includes 1,049,192 reusing papers involving 10,000 randomly sampled new word combinations reused at least 10 times. For *New Phrase Combination*, the sample includes 410,406 reusing papers involving 10,000 randomly sampled new phrase combinations reused at least 10 times. All models include fixed effects for the subfield of both the pioneering and reusing papers, as well as publication year fixed effects for the pioneering paper. Additional controls for both the pioneering and reusing papers include whether an abstract is available, text length (number of unique words and phrases in the title and abstract), and the number of unique papers and journals cited.

**Appendix E. Literature Review Papers**

We collect a large sample of literature review papers and a matched control sample of original non-review papers. Our assumption is that review papers primarily summarize prior scientific findings rather than pioneering new insights. Therefore, we expect review papers to be less likely to introduce new scientific ideas compared to the matched control papers (Wu, Wang, & Evans, 2019). Put differently, we assume control papers are more likely to pioneer new insights than review papers. However, most scientific papers are not highly original or novel, often representing only incremental advances over the existing body of knowledge (e.g., Uzzi et al., 2013; Wang, Veugelers, & Stephan, 2017). As a result, unlike Nobel Prize papers, which likely present highly original and novel ideas, we expect the average control paper in our sample to score relatively low on novelty. Consequently, it may be more challenging for the metrics to distinguish between review papers and control papers compared to distinguishing Nobel Prize papers from controls, making review papers a less convincing validation test. Additionally, unlike Nobel Prize papers, the average control paper in our sample is unlikely to introduce new ideas with significant impact. Therefore, our validation will focus on ex ante metrics that capture novelty at the time of publication. While this approach has some drawbacks compared to the Nobel Prize validation, its advantage lies in the ability to collect a large, diverse sample of review papers across all fields of study, rather than relying on a smaller sample of Nobel Prize papers that, while likely covering very original scientific insights, are also selected based on their significant impact on later work.

*Data*

We identify all scientific papers in history that include both the words "literature" and "review" in their titles. We then apply the same case-control study design as for the Nobel Prize papers, matching each review paper to one control paper –a regular paper that does not mention "literature" or "review" in its title or abstract but was published in the same year, journal, and subfield as the review paper. In cases where multiple matching control papers are available, we randomly select one. Matching on journal, year, and subfield controls for differences across disciplines and time. To account for journals that exclusively publish review papers, we exclude any papers (both review and control) published in journals with "literature" or "review" in their titles (e.g., Journal of Economic Literature).[28] Given our research design, it is preferable to eliminate false negative review papers rather than cover the entire population of review papers. Finally, we test the ability of all metrics to distinguish between review and control papers. Our sample consists of 34,428 review papers published between 1901 and 2010, along with 34,428 matched control papers.

*Results*

Table E.1 presents descriptive statistics for review versus control papers. All text metrics indicate that review papers exhibit significantly lower novelty compared to control papers, as confirmed by a Mann-Whitney test with p=0.000. By contrast, traditional citation-based measures of novelty (*Wang* and *Uzzi*) suggest that review papers are more novel at the time of publication. This discrepancy is likely due to the broader and more interdisciplinary references covered by review papers, leading to more novel and atypical combinations of cited journals, thus inflating their novelty score. For this reason, review papers are typically excluded when assessing novelty using these measures (e.g., Wang, Veugelers, & Stephan, 2017). Consistent with the Nobel Prize validation, *New Phrase* performs best in distinguishing review papers from matched control papers at the time of publication, as shown by the highest test statistic in the Mann-Whitney test.

　　　Table E.2 displays the results of logit regressions with a binary indicator for review papers as the outcome. The regression results align closely with the descriptive statistics. All new text metrics indicate that

---

[28] We manually checked a random sample of 100 papers (50 papers with "literature" and "review" in their title and 50 control papers) and confirmed that all papers classified as review papers were literature reviews, and all control papers were not.

review papers have significantly lower novelty than control papers. However, as expected, the overall predictive power of the metrics to distinguish review papers from regular papers is lower than for Nobel Prize papers, and the differences between the metrics are minor. Similarly, the differences in performance between the models including text metrics and the baseline model including only controls are small. In contrast to the findings for Nobel Prize papers, *Semantic Distance* becomes significant, indicating that review papers are indeed less novel. Nevertheless, consistent with the Nobel Prize validation, *New Phrase* remains the best-performing metric overall, with a precision of 60.29%, recall of 63.40%, and AUC of 0.6601. A one standard deviation increase in *New Phrase* reduces the likelihood of being a review paper by 8.62%.

**Table E.1: Descriptive Statistics Literature Review versus Control Papers**

| | Mean | Stdev | Min | p25 | p50 | p75 | p95 | p99 | Max | Mean | Stdev | Min | p25 | p50 | p75 | p95 | p99 | Max | Z | p-value |
|---|---|---|---|---|---|---|---|---|---|---|---|---|---|---|---|---|---|---|---|---|
| | Literature Review papers (n=34,428) | | | | | | | | | Control papers (n=34,428) | | | | | | | | | | | |
| **Panel A: Ex ante** | | | | | | | | | | | | | | | | | | | | | |
| New Word (Binary) | 0.045 | 0.207 | 0.000 | 0.000 | 0.000 | 0.000 | 0.000 | 1.000 | 1.000 | 0.087 | 0.281 | 0.000 | 0.000 | 0.000 | 0.000 | 1.000 | 1.000 | 1.000 | 22.021 | 0.0000*** |
| New Word | 0.034 | 0.161 | 0.000 | 0.000 | 0.000 | 0.000 | 0.000 | 0.693 | 2.565 | 0.069 | 0.236 | 0.000 | 0.000 | 0.000 | 0.000 | 0.693 | 1.099 | 3.555 | 22.122 | 0.0000*** |
| New Phrase (Binary) | 0.257 | 0.437 | 0.000 | 0.000 | 0.000 | 1.000 | 1.000 | 1.000 | 1.000 | 0.421 | 0.494 | 0.000 | 0.000 | 0.000 | 1.000 | 1.000 | 1.000 | 1.000 | 45.264 | 0.0000*** |
| New Phrase | 0.213 | 0.382 | 0.000 | 0.000 | 0.000 | 0.693 | 1.099 | 1.386 | 2.708 | 0.399 | 0.519 | 0.000 | 0.000 | 0.000 | 0.693 | 1.386 | 1.792 | 3.466 | 49.733 | 0.0000*** |
| New Word Combination (Binary) | 0.604 | 0.489 | 0.000 | 0.000 | 1.000 | 1.000 | 1.000 | 1.000 | 1.000 | 0.688 | 0.463 | 0.000 | 0.000 | 1.000 | 1.000 | 1.000 | 1.000 | 1.000 | 22.849 | 0.0000*** |
| New Word Combination | 1.346 | 1.392 | 0.000 | 0.000 | 1.099 | 2.398 | 3.807 | 4.745 | 8.506 | 1.842 | 1.648 | 0.000 | 0.000 | 1.792 | 3.135 | 4.554 | 5.787 | 9.925 | 38.980 | 0.0000*** |
| New Phrase Combination (Binary) | 0.669 | 0.470 | 0.000 | 0.000 | 1.000 | 1.000 | 1.000 | 1.000 | 1.000 | 0.729 | 0.444 | 0.000 | 0.000 | 1.000 | 1.000 | 1.000 | 1.000 | 1.000 | 17.212 | 0.0000*** |
| New Phrase Combination | 1.528 | 1.390 | 0.000 | 0.000 | 1.386 | 2.639 | 3.829 | 4.625 | 9.003 | 1.956 | 1.601 | 0.000 | 0.000 | 2.079 | 3.258 | 4.431 | 5.421 | 9.638 | 35.244 | 0.0000*** |
| Semantic Distance | 0.118 | 0.044 | 0.010 | 0.088 | 0.112 | 0.141 | 0.198 | 0.247 | 0.446 | 0.124 | 0.043 | 0.010 | 0.095 | 0.119 | 0.148 | 0.203 | 0.252 | 0.379 | 22.379 | 0.0000*** |
| Wang | 0.256 | 0.609 | 0.000 | 0.000 | 0.000 | 0.000 | 1.638 | 2.843 | 6.284 | 0.125 | 0.391 | 0.000 | 0.000 | 0.000 | 0.000 | 0.978 | 1.933 | 4.865 | -31.877 | 0.0000*** |
| Uzzi | 4.082 | 0.285 | 1.605 | 4.013 | 4.043 | 4.079 | 4.415 | 5.469 | 8.375 | 4.119 | 0.431 | -0.012 | 4.014 | 4.043 | 4.092 | 4.785 | 6.219 | 8.682 | 3.947 | 0.0000*** |

*Notes*: n=68,856 papers of which 34,428 literature review papers and 34,428 matched control papers published between 1901 and 2010. Each literature review paper is matched to one randomly selected control paper published in the same year, journal and subfield. p25, p50 p75, p95 and p99 are respectively the 25th, 50th, 75th, 95th and the 99th percentile. All measures except binary indicators, *Semantic Distance* and *CD* are log transformed after adding 1 for measures with 0 values. Z values are test statistics from the Mann-Whitney test. *** p<0.01, ** p<0.05, * p<0.10

**Table E.2: Likelihood of Literature Review Paper with *ex ante* Metrics**

| | Text Metrics | | | | | Traditional Metrics | | Text Metrics Combined | All Metrics Combined |
|---|---|---|---|---|---|---|---|---|---|
| **Panel A: Ex ante** | | | | | | | | | |
| | (1) | (2) | (3) | (4) | (5) | (6) | (7) | (10) | (11) |
| New Word | -0.508*** | | | | | | | 0.002 | 0.002 |
| | (0.043) | | | | | | | (0.046) | (0.046) |
| New Phrase | | -0.805*** | | | | | | -0.722*** | -0.723*** |
| | | (0.021) | | | | | | (0.022) | (0.023) |
| New Word Combination | | | -0.188*** | | | | | -0.084*** | -0.089*** |
| | | | (0.009) | | | | | (0.010) | (0.010) |
| New Phrase Combination | | | | -0.197*** | | | | -0.075*** | -0.077*** |
| | | | | (0.013) | | | | (0.015) | (0.015) |
| Semantic Distance | | | | | -4.474*** | | | -4.409*** | -4.829*** |
| | | | | | (0.228) | | | (0.231) | (0.234) |
| Wang | | | | | | 0.402*** | | | 0.418*** |
| | | | | | | (0.019) | | | (0.019) |
| Uzzi | | | | | | | -0.326*** | | -0.303*** |
| | | | | | | | (0.023) | | (0.023) |
| Log-Likelihood | -45,541 | -44,860 | -45,391 | -45,503 | -45,412 | -45,376 | -45,480 | -44,607 | -44,220 |
| Pseudo-R² | 0.046 | 0.060 | 0.046 | 0.046 | 0.048 | 0.046 | 0.046 | 0.065 | 0.073 |
|    Precision (%) | 60.72 | 60.29 | 60.64 | 60.65 | 61.68 | 61.80 | 61.04 | 61.20 | 62.02 |
|    Recall (%) | 61.20 | 63.40 | 60.16 | 60.16 | 61.03 | 57.48 | 60.17 | 64.14 | 63.57 |
|    AUC | 0.6441 | 0.6601 | 0.6466 | 0.6433 | 0.6505 | 0.6467 | 0.6447 | 0.6670 | 0.6750 |
|    Marginal Effects (%) | -2.42 | -8.62 | -6.79 | -6.99 | -4.57 | 4.85 | -2.80 | | |

*Notes*: Logit regressions, robust standard errors in parentheses. n=68,856 papers of which 34,428 literature review papers and 34,428 matched control papers published between 1901 and 2010. Each literature review paper is matched to one randomly selected control paper published in the same year, journal and subfield. All measures except *Semantic Distance* are log transformed after adding 1 for measures with 0 values. All models include publication year and subfield of study fixed effects, and additionally control for whether the paper has an abstract available, text length (number of unique words and phrases in the title and abstract of the paper), and the number of unique papers and journals cited by the focal paper. AUC is area under the ROC curve. Marginal effects are calculated as the % increase in the likelihood of being a literature review paper associated with an increase in the metric with one standard deviation. Baseline model (only with controls) performances (not shown in table): Precision=60.91, recall=59.88, AUC=0.6412. *** p<0.01, ** p<0.05, * p<0.10

## Appendix F. Publication Timing of Intellectual Neighbors

This validation exercise tests the hypothesis that more novel papers tend to have a greater proportion of their closest intellectual neighbors—defined as the most similar papers—published after them, compared to less novel papers. In other words, if a paper introduces a new scientific idea, it is expected that related research will be published after that paper, indicating that the paper is ahead of its intellectual peers.[29]

### *Data*

We leverage the PubMed Related Citations Algorithm (PMRCA) to identify the most similar articles for each paper in PubMed based on title, abstract, and MeSH terms (Lin & Wilbur, 2007). MeSH terms, curated by the National Library of Medicine, provide detailed classification of the biomedical literature. We identified 25,097,834 papers present in both OpenAlex and PubMed, using PubMed identifiers provided by OpenAlex. For each paper, we gathered its five closest intellectual neighbors and calculated the distance in days between the focal paper and its related articles.[30] We then computed the *After Share*, defined as the proportion of intellectual neighbors published after the focal paper.

### *Results*

Using fractional logit models (Papke & Wooldridge, 1996), we analyzed how the different text-based and citation-based novelty metrics influence the share of a paper's intellectual neighbors published after it. The analysis, restricted to papers published between 1901 and 2010 (n=11,543,342), controls for text length, abstract availability, number of references, number of unique journals referenced, and includes fixed effects for publication year and subfield. On average, 61.5% of a paper's closest intellectual neighbors are published after it, with a median of 60%. The fractional logit results, presented in Table F.1, indicate that higher novelty is associated with a larger share of intellectual neighbors being published later. Among all metrics, *New Phrase Combination* exhibits the strongest effect, with a one-standard deviation increase resulting in a 12.72% rise in the share of later-published neighbors.

### Table F.1: Share of Intellectual Neighbors Published after Focal Paper

| | Text Metrics | | | | | Traditional Metrics | | Text Metrics Combined | All Metrics Combined |
|---|---|---|---|---|---|---|---|---|---|
| **Panel A: Ex ante** | | | | | | | | | |
| | (1) | (2) | (3) | (4) | (5) | (6) | (7) | (8) | (9) |
| New Word | 0.159*** | | | | | | | 0.032*** | 0.033*** |
| | (0.002) | | | | | | | (0.002) | (0.002) |
| New Phrase | | 0.262*** | | | | | | 0.160*** | 0.160*** |
| | | (0.001) | | | | | | (0.001) | (0.001) |
| New Word Combination | | | 0.111*** | | | | | 0.009*** | 0.009*** |
| | | | (0.000) | | | | | (0.000) | (0.000) |
| New Phrase Combination | | | | 0.374*** | | | | 0.361*** | 0.359*** |
| | | | | (0.001) | | | | (0.001) | (0.001) |
| Semantic Distance | | | | | 1.449*** | | | 2.035*** | 1.989*** |
| | | | | | (0.011) | | | (0.011) | (0.011) |
| Wang | | | | | | 0.137*** | | | 0.109*** |
| | | | | | | (0.001) | | | (0.001) |
| Uzzi | | | | | | | 0.001 | | 0.033*** |
| | | | | | | | (0.003) | | (0.003) |
| Pseudo $R^2$ | 0.0290 | 0.0308 | 0.0304 | 0.0365 | 0.0292 | 0.0291 | 0.0287 | 0.0384 | 0.0387 |
| Marginal Effects (%) | 0.93 | 2.89 | 4.03 | 12.72 | 1.36 | 1.19 | 0.01 | | |

*Notes:* Fractional logit models, robust standard errors in parentheses. The dependent variable is the share of later-published top 5 intellectual neighbors. The sample includes papers published between 1901 and 2010 (n=11,543,342). All measures except *Semantic Distance* and *CD* are log transformed after adding 1 for measures with 0 values. All models include publication year and subfield of study fixed effects, and additionally control for whether the paper has an abstract available, text length (number of unique words and phrases in the title and abstract of the paper), and the number of unique papers and journals cited by the focal paper. The marginal effects represent the percentage change in the proportion of intellectual neighbors published after the focal paper with a one-standard-deviation increase in the metric. *** p<0.01, ** p<0.05, * p<0.10.

---

[29] We thank Pierre Azoulay for this suggestion.
[30] We used the Entrez Programming Utilities (E-utilities) API, accessed between 2024.06.01 and 2024.06.12.

## Appendix G. Likelihood of Language that denotes Novelty

We investigate whether papers introducing new scientific ideas are more likely to use language that denotes novelty. Following the approach developed by Leahey et al. (2023), we construct a binary indicator for each paper, indicating whether the (not processed) title or abstract includes at least one word that denotes novelty, such as "discover," "introduce," "novel," and "innovative."[31] Among all papers published between 1901 and 2023 (n=75,295,921), 21% of papers include at least one word denoting novelty in the title or abstract. Using OLS regressions, we examine the relationship between the novelty metrics and the use of language that denotes novelty, controlling for text length, abstract availability, number of references, number of unique journals referenced, and includes fixed effects for publication year and subfield. The results, shown in Table G.1, indicates that higher novelty is associated with higher likelihood of using language that denotes novelty. Among all metrics, *New Word Combination* has the strongest effect: a one-standard deviation increase in this metric is linked to a 1.57% higher probability of novelty language.

### Table G.1: Likelihood of Language that Denotes Novelty

| | Text Metrics | | | | | Traditional Metrics | | Text Metrics Combined | All Metrics Combined |
|---|---|---|---|---|---|---|---|---|---|
| **Panel A: Ex ante** | | | | | | | | | |
| | (1) | (2) | (3) | (4) | (5) | (6) | (7) | (8) | (9) |
| New Word | 0.040*** | | | | | | | 0.020*** | 0.020*** |
| | (0.000) | | | | | | | (0.000) | (0.000) |
| New Phrase | | 0.023*** | | | | | | 0.015*** | 0.015*** |
| | | (0.000) | | | | | | (0.000) | (0.000) |
| New Word Combination | | | 0.011*** | | | | | 0.010*** | 0.010*** |
| | | | (0.000) | | | | | (0.000) | (0.000) |
| New Phrase Combination | | | | 0.001*** | | | | -0.007*** | -0.007*** |
| | | | | (0.000) | | | | (0.000) | (0.000) |
| Semantic Distance | | | | | 0.095*** | | | 0.072*** | 0.070*** |
| | | | | | (0.001) | | | (0.001) | (0.001) |
| Wang | | | | | | 0.009*** | | | 0.008*** |
| | | | | | | (0.000) | | | (0.000) |
| Uzzi | | | | | | | -0.005*** | | -0.005*** |
| | | | | | | | (0.000) | | (0.000) |
| $R^2$ | 0.1181 | 0.1182 | 0.1183 | 0.1178 | 0.1179 | 0.1178 | 0.1178 | 0.1188 | 0.1188 |
| Marginal Effects (%) | 0.81 | 0.92 | 1.57 | 0.11 | 0.44 | 0.27 | -0.07 | | |

*Notes:* OLS models, robust standard errors in parentheses. The dependent variable is a binary indicator of whether a paper contains language that denotes novelty in the (not processed) title or abstract (mean=0.212; std=0.409). The sample includes papers published between 1901 and 2023 (n=75,295,921). All measures except *Semantic Distance* and *CD* are log transformed after adding 1 for measures with 0 values. All models include publication year and subfield of study fixed effects, and additionally control for whether the paper has an abstract available, text length (number of unique processed words and phrases in the title and abstract of the paper), and the number of unique papers and journals cited by the focal paper. Marginal effects represent the percentage increase in the probability of containing language that denotes novelty associated with a one-standard-deviation increase in the metric. *** p<0.01, ** p<0.05, * p<0.10

## Appendix H. New Scientific Ideas and Citation Outcomes

We investigate whether papers introducing new scientific ideas tend to receive more citations and have a higher likelihood of becoming highly cited.

### Data

We calculate the total number of citations each paper in the OpenAlex dataset received from subsequent papers up to 2023 (mean = 29.19, standard deviation = 154.11, median = 6) and analyze how various metrics influence citation outcomes using two complementary approaches.[32] First, we employ Poisson quasi-maximum likelihood models to estimate the effects of these metrics on citation counts. Second, we define top-cited papers as those receiving more citations than the 95th percentile of all papers published in the same subfield and year and, alternatively, as those exceeding the 99th percentile. We then use linear probability models to assess how each metric affects the likelihood of a paper being top-cited. To quantify and visualize the relationship between novelty metrics and citation outcomes, we create indicator variables based on each paper's novelty metric value relative to the distribution of this metric among papers from the same subfield and year. Specifically, we divide the top 10% of each metric's distribution into five 2-percentile buckets (p90–p92, p92–p94, p94–p96, p96–p98, and p98–p100), while treating the p0–p90 range as a single baseline category. This approach provides greater resolution within the top 10%, where variation among papers is most pronounced, while consolidating the lower 90% into a single category to account for its limited variability.[33] For example, 93% of all papers introduce no new words, and 92% of all papers have the minimum value for *Wang*. To avoid truncation bias, we focus on the complete set of papers published between 1901 and 2010 (n=37,154,406). All estimations include the same control variables as in previous analyses: a binary indicator for whether the paper has an abstract, the number of unique words and phrases in the title and abstract, the number of unique cited papers and journals, and fixed effects for the subfield of study and publication year.

---

[31] The words include "advancement", "breakthrough", "create", "discover", "demonstrate", "develop", "devise", "innovative", "introduce", "new", "novel", "original", "propose", "unknown", "unique", and "unrecognized." To minimize false positives, we follow the detailed guidelines specified in Appendix B of Leahey et al. (2023). For instance, the word "new" is not considered when it appears in contexts such as proper nouns (e.g., "New York"), institutional terms (e.g., "new journal"), or negated expressions (e.g., "not new"); similarly, the stem "origin" is not considered when followed by terms like "edition" or "publication," and the word "first" is only considered when directly followed by words such as "principles" or "experiment."

[32] As of January 2024, OpenAlex aggregates citation data from sources such as Crossref, PubMed, Unpaywall, the Directory of Open Access Journals, arXiv, Zenodo, and the International Standard Serial Number Centre. It builds on the Microsoft Academic Graph and integrates current data from trusted scholarly sources, ensuring comprehensive and reliable coverage.

***Results***

Figures H.1, H.2, and Figure 2 in the main text illustrate the predicted number of citations and the predicted likelihood of a paper being top-cited, measured as being above the 95th percentile and above the 99th percentile, respectively, across different percentile ranges of each novelty metric. Each figure is divided into 12 panels, with each panel corresponding to a specific metric.

Our findings demonstrate that text-based metrics capturing a paper's novelty at the time of publication (i.e., ex ante) are generally more effective predictors of both citation counts and the likelihood of being top-cited than traditional citation-based novelty metrics (*Uzzi* and *Wang*). These results underscore the critical role of introducing new scientific ideas as a driver of scientific progress. However, there are notable exceptions. *Semantic Distance* negatively impacts citation outcomes, suggesting that papers with higher semantic distance from prior work may struggle to gain recognition. Moreover, *New Word* is a weaker predictor of citation outcomes when compared to the other text-based metrics.

Reassuringly, text-based metrics that measure the reuse of new scientific ideas in subsequent publications (i.e., ex post) generally outperform ex ante metrics in predicting citation outcomes. This highlights the effectiveness of reuse-based metrics in capturing the broader influence and impact of new scientific ideas on subsequent literature, as reflected in citation counts. Notably, *CD* strongly correlates with overall citation counts.

---

[33] In cases where multiple 2-percentile buckets share identical values, papers are assigned to the middle bucket within the overlapping range. To ensure robustness, we tested several alternative approaches: (1) assigning observations to the first bucket within the overlapping range instead of the middle bucket, (2) using 1-percentile buckets instead of 2-percentile buckets, (3) treating the novelty metric as a continuous or count variable rather than converting it into multiple binary indicators, and (4) categorizing papers based on whether they fall above the 99th percentile for a specific metric within the same subfield and year versus not. Our results remained robust across all these alternative specifications, underscoring the consistency of our findings.

## Figure H.1: Predicted Number of Citations

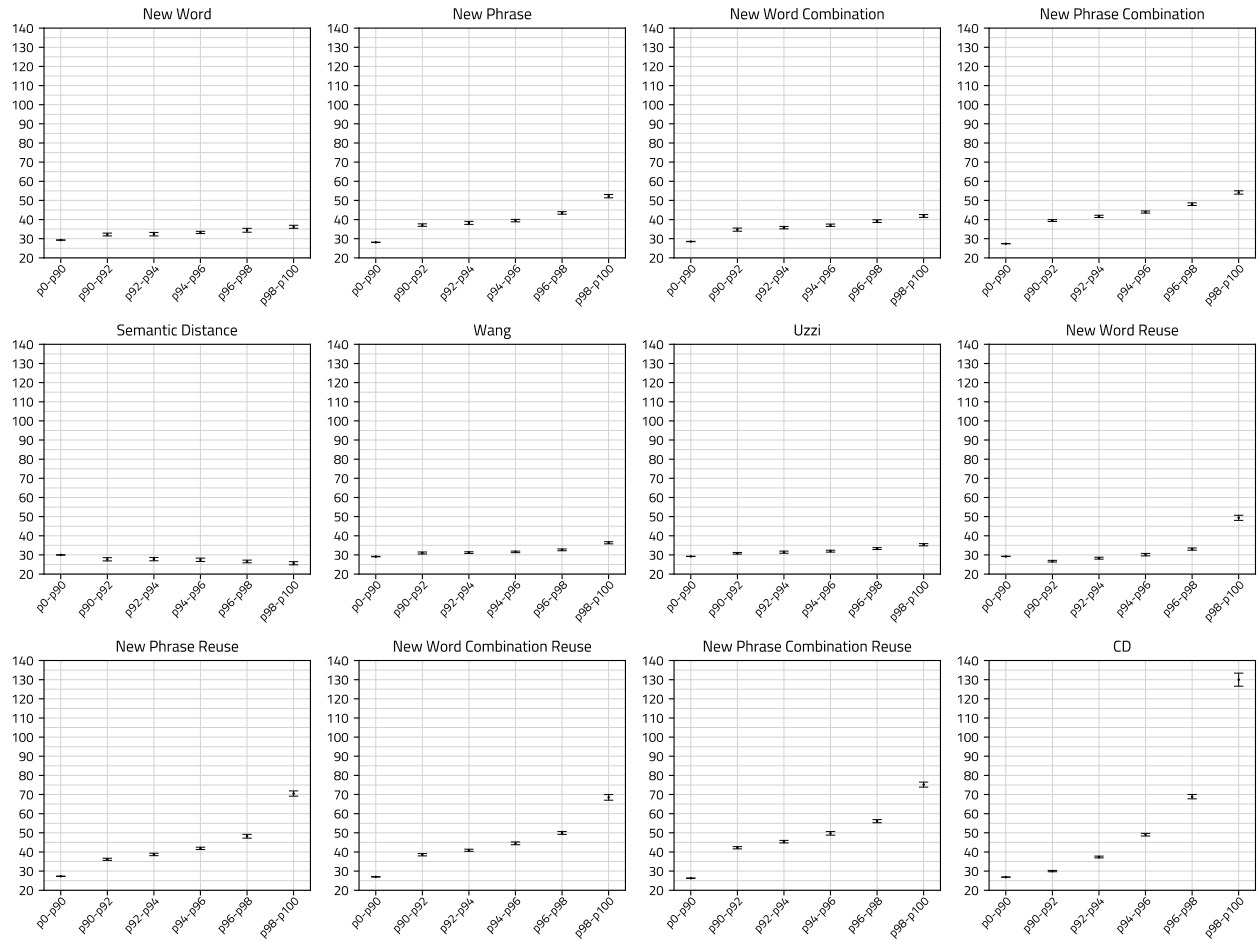

*Notes:* Figure H.1 plots the predicted number of citations, estimated using 12 separate quasi-maximum likelihood Poisson models. Each model includes indicator variables for whether a paper's metric value, within its subfield and year, falls into the 0th–90th percentile (reference category) or one of five 2-percentile ranges from the 90th to 100th percentile. For consistency in comparisons, *Uzzi* is inverted (e.g., p98–p100 corresponds to p0–p2), ensuring higher novelty aligns with higher percentile values. The predicted number of citations are shown with 99.999% confidence intervals. Control variables in the models include whether the paper has an abstract, the number of unique words and phrases in title and abstract, the number of unique cited papers and journals, and fixed effects for subfield and year of publication. The figure is organized into 12 panels, each representing a different metric. The analysis includes all papers from the OpenAlex database published between 1901 and 2010 (n=37,154,406).

**Figure H.2: Predicted Probability of Paper Being Top Cited (above 95th Percentile of Citations)**

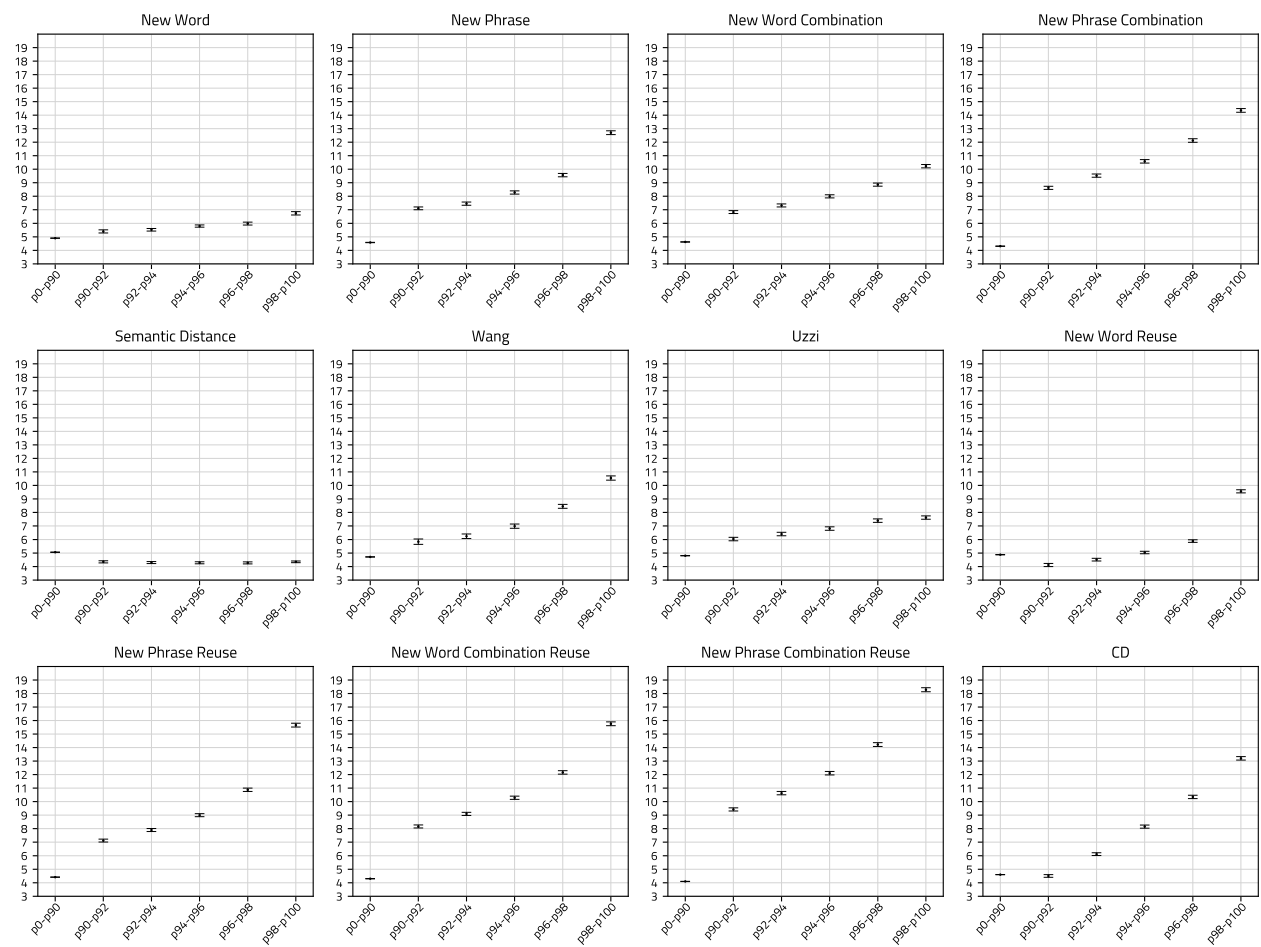

*Notes:* Figure H.2 plots the predicted probability (in %) of a paper being among the top 5% most-cited (above the 95th percentile of received citations within its sub-field and year), estimated using 12 separate linear probability models. Each model includes indicator variables for whether a paper's metric value, within its subfield and year, falls into the 0th–90th percentile (reference category) or one of five 2-percentile ranges from the 90th to 100th percentile. For consistency in comparisons, *Uzzi* is inverted (e.g., p98–p100 corresponds to p0–p2), ensuring higher novelty aligns with higher percentile values. The predicted probabilities are shown with 99.999% confidence intervals. Control variables in the models include whether the paper has an abstract, the number of unique words and phrases in title and abstract, the number of unique cited papers and journals, and fixed effects for subfield and year of publication. The figure is organized into 12 panels, each representing a different metric. The analysis includes all papers from the OpenAlex database published between 1901 and 2010 (n=37,154,406).

# Appendix I. Limitations of Text Metrics and Suggestions for Future work

First, unlike citations, text data is unstructured and often messy, particularly for older papers where titles and abstracts are sometimes recovered through optical character recognition (OCR), potentially introducing errors and bias. Since we rely on OpenAlex, the quality of our metrics and findings is greatly influenced by the accuracy of this source data. We encourage others to replicate our metrics and findings using other publication databases, such as Web of Science or Scopus, which may be better curated. However, the key drawback of these alternatives is that they typically do not offer open access.

Second, while lemmatization helps standardize different spellings of the same word, it does not fully address spelling errors, synonyms (different words with the same meaning), or homonyms (same word with different meanings). Although our use of paper-level embeddings accounts for synonyms, homonyms, and the context of words in a publication, the embedding-based metric developed in this paper did not perform well in identifying novelty. This outcome may be surprising given that we used SPECTER, currently one of the best-performing document-level embeddings for scientific papers. However, this metric seems better suited for measuring similarity between publication documents rather than identifying new scientific ideas (Beltagy et al., 2019; Cohan et al., 2020; Erhardt et al., 2022).

Another limitation is that new scientific ideas and concepts may only receive a specific label after the pioneering paper is published, or the pioneering paper may not use the label in its title or abstract. As a result, we might misattribute the introduction of a new idea to a later paper rather than the true pioneer. Additionally, due to open access restrictions, we were limited to analyzing titles and abstracts, while other sections of the paper—such as the introduction, conclusion, or the description of the contribution relative to prior work—likely contain valuable information for identifying new scientific ideas and their impact.

Beyond limitations related to text, another issue concerns the sample of publications included in our analysis. Ideally, we would include every scientific publication in history to trace the scientific frontier over time and accurately assign new insights to the right papers. We relied on OpenAlex, the largest open-access database of scientific publications. However, OpenAlex likely does not contain every paper published in history and may be messier and less reliable than other databases like Web of Science or Scopus. An important limitation is that we had to exclude a significant number of publications due to potential issues in the recorded data.

We recommend that future users further clean measurement errors, experiment with alternative metrics, control for the availability of abstracts, control for paper text length, and account for differences across fields of study and over time. A promising avenue for future research is to employ more advanced NLP tools, such as Named Entity Recognition (NER), to extract the most coherent and contextually relevant phrases from publications. This approach could help better define and operationalize the concept of a new scientific idea. Leveraging language models to train NER could improve the identification of these ideas and potentially classify them into coherent categories, such as research questions, methods, data, findings, and contributions. This would enable a more nuanced identification of novel contributions and provide richer data to study scientific progress (Luo et al., 2022; Cheng et al., 2023; Leahey et al., 2023; Shi & Evans, 2023).